\documentclass[12pt]{article}
\usepackage{cite}
\usepackage{a4wide}
\usepackage{graphicx}
\usepackage{amsfonts}
\usepackage{amstext}
\usepackage{amsmath}
\usepackage{amssymb}
\usepackage{amsthm}
\usepackage[mathscr]{eucal}
\makeatletter
 
 \@addtoreset{equation}{section}
 \newtheorem{theorem}{Theorem}
\newtheorem{proposition}[theorem]{Proposition}

\newtheorem{corollary}[theorem]{Corollary}
  \@addtoreset{theorem}{section}

\makeatother
\def\a{\alpha}

\def\C{\mathbb{C}}

\def\e{\epsilon}
\def\ve{\varepsilon}

\def\g{\gamma}

\def\i{\infty}

\def\l{\lambda}

\def\N{\mathbb{N}}
\def\o{\omega}

\def\1{\bf{1}}

\def\sgn{{\rm{sgn}}}

\def\th{\theta}
\def\Tr{{\rm Tr}}

\def\Z{\mathbb{Z}}

\def\erfc{\text{erfc}}

\begin{document}
\title{Distribution of a tagged particle position in the one-dimensional symmetric simple exclusion process
        with two-sided Bernoulli initial condition}

\author{Takashi Imamura
\footnote { Department of mathematics and informatics, 
Chiba University,~E-mail: imamura@math.s.chiba-u.ac.jp}
, Kirone Mallick
\footnote { Institut de Physique Th\'eorique, 
CEA Saclay and URA 2306,
~E-mail: kirone.mallick@ipht.fr}
, Tomohiro Sasamoto
\footnote { Department of physics,
Tokyo Institute of Technology,~E-mail: sasamoto@phys.titech.ac.jp}}
\maketitle

\begin{abstract}
For the two-sided Bernoulli initial condition with density $\rho_-$ (resp. $\rho_+$) to the 
left (resp. to the right), we study the distribution of a tagged particle in the one dimensional symmetric 
simple exclusion process. We obtain a formula for the moment generating
function of the associated current in terms of 
a Fredholm determinant. Our arguments are based on a combination of techniques from integrable probability 
which have been developed recently for studying the asymmetric exclusion process and a subsequent intricate 
symmetric limit. An expression for the large deviation function of the tagged 
particle position is obtained, including the case of
the stationary measure with uniform density $\rho$. 


\end{abstract}

\section{Introduction and results} 

\subsection{The exclusion process and a tagged particle}

  The asymmetric simple exclusion process (ASEP) is a continuous time interacting particles
  Markov process $\eta(t)=\{\eta_x(t),x\in\mathbb{Z}\}$, in which each particle is located on a
  discrete site labeled by an integer $x \in {\mathbb Z}$ and can hop to its  right or left  nearest 
  neighboring site with rate $p$ or $q$ respectively, where $0\leq p,q\leq 1, p+q>0$.
  Due to the volume exclusion, jumps to an occupied site are forbidden.  
  The state of the ASEP, $\eta=\{\eta_x\}_{x\in\mathbb{Z}}$, is a collection of $\eta_x$ which is 
   the occupation at $x$,  i.e., $\eta_x=1$(resp. 0) when the site $x$ is occupied (resp. empty).  
   Formally the ASEP is defined through the generator,  
  \begin{equation}
   L f = \sum_{x\in\mathbb{Z}} (p\eta_x(1-\eta_{x+1})+q(1-\eta_x)\eta_{x+1})[f(\eta^{x,x+1})-f(\eta)],
  \end{equation}
  where 
   $\eta^{x,x+1}$ denotes the state with 
   $\eta_x,\eta_{x+1}$ swapped in $\eta$, and $f$ is a cylinder function. 
   For  the  precise construction  of the process,  see \cite{Liggett1985,Liggett1999}.
    The asymmetry  parameter is defined  as $\tau = p/q$ and in this paper we follow the convention of 
   \cite{TW2008a,BCS2014} in which particles hop preferably towards the left, i.e., $0\leq \tau\leq 1$.
   Accordingly, a current flowing from right to left will be  counted positively. 

  The symmetric simple exclusion process (SEP) corresponds to the case
  where the jumps are isotropic; the 
  hopping rates are equal and set to  $p=q=1$. 
 In this work, we  shall  focus  on the motion of a
  tagged particle (or tracer) in
  the SEP. Because of the exclusion condition, particles move and
  conserve their ordering. The SEP is therefore a pristine model of
  {\it single file} diffusion \cite{PMRichards,Pincus,Bechinger} and has a 
   lot of applications in physics, see for instance  \cite{KMSprl,Ryabov}
   for recent references. 
  
  The long time behavior of the tracer $X_t$ has been the subject
   of many studies since Spitzer's original paper  \cite{Spitzer1970}. 
  In particular,  for  a  system initially at equilibrium with  uniform density $\rho$,
  Arratia \cite{Arratia1983}, Rost and Vares \cite{RostVares} and De Masi and Ferrari \cite{DeMasiFerrari}
   proved that the  variance of the position $X_t$ at time $t$ of the tracer grows 
  subdiffusively with time as $t^{1/2}$ and further that 
  the rescaled variable  $t^{-1/4} X_t$ converges to a
   Gaussian with variance $\sigma_X^2 = \frac{2 ( 1 -
     \rho)}{\rho\sqrt{\pi}}$. Moreover,  Peligrad and
    Sethuraman  \cite{SethuramPeligrad} have established the generalization 
   (see for instance \cite{SpohnBook},  Chapter 6, Conjecture 6.5 page 294)
    that the rescaled process weakly converges to a
   fractional Brownian motion ${\mathbb B}_{1/4}(t)$ with Hurst parameter 1/4,  
   $$\sigma_X^{-1} \lambda^{-1/4} X_{\lambda t}  \Rightarrow  {\mathbb B}_{1/4}(t).$$

    For  an  initial setting  out of equilibrium,
   that corresponds to a non-uniform initial distribution of  the
   particles, Jara and Landim \cite{JaraLandim} 
  have obtained  laws of large numbers and
  central limit theorems for local equilibrium initial settings. The former reads 
\begin{equation}
\frac{X_t}{\sqrt{4t}} \to 
 -\xi_0
\label{AveragePosition}
\end{equation}
as $t\to\infty$ in probability with the value of $\xi_0$ being given by
\begin{equation}
\int_0^{\infty} (\rho(x,t)-\rho(x,0)) dx = \int_0^{-\xi_0} \rho(x,t) dx.
\label{rhoxi0}
\end{equation}
Here $\rho(x,t)$ is the solution to the hydrodynamic equation, 
which for the case of SEP is simply given by the diffusion equation
(see for instance \cite{KipnisLandim1999}).

\subsection{Large deviation for a tagged particle} 
     More recently, Sethuraman and Varadhan \cite{Sethuraman2013}  have
 established the large deviation principle for a tagged particle in SEP. 
(For general information about large deviation theory, see for instance 
\cite{DemboZeitouni, denHollander2000, Varadhan1984}.)
 They  considered  two classes of initial conditions -- either deterministic
  configurations or local equilibrium measures that interpolate 
   between a density $\rho_{-}$ when $x \to -\infty$ and 
   $\rho_{+}$ when $x \to +\infty$ -- and 
    proved that there exists  a large deviation (or good rate)  function $\phi(\xi)$, i.e., 
    a function on $\mathbb{R}$
     such that for each $a\in[0,\infty)$,  the set $\{ \xi: \phi(\xi) \leq a \}$   is 
    closed and compact,  and the following property holds
\begin{align}
 &\quad -\inf_{\xi\in\bar{U}} \phi(\xi) \geq \limsup_{t\to\infty} t^{-1/2} \text{Prob}\left( -\frac{X_t}{\sqrt{4 t}} \in U \right) \notag\\
 &\geq  \liminf_{t\to\infty} t^{-1/2} \text{Prob}\left( -\frac{X_t}{\sqrt{4 t}} \in U\right) \geq -\inf_{\xi\in U^{\circ}} \phi(\xi),
\end{align}
where $U$ is a Borel set of $\mathbb{R}$, $U^{\circ}$ denotes its interior, $\bar{U}$ its closure. 
Since our main interest in the paper is the explicit calculation of the rate function, 
we will write this statement more heuristically as 
   \begin{equation}
  \text{Prob}\left( \frac{X_t}{\sqrt{4 t}} = -\xi  \right)
   \simeq  \exp[-\sqrt{t} \phi(\xi)] . 
 \label{def:phi}
  \end{equation}
Here the left hand side is an abuse of notation for the probability 
$\text{Prob}\left( -\frac{X_t}{\sqrt{4 t}} \in (\xi,\xi+d\xi]  \right)$ and 
equalities with the symbol $\simeq$  have to be 
 understood at the level of dominant exponential terms, {\it e.g.}, 
  (\ref{def:phi}) means $\lim_{t \to \infty} \frac{-1}{\sqrt{t}}
 \log  \Big[ \text{Prob}\left( -\frac{X_t}{\sqrt{4 t}} \in (\xi,\xi+d\xi]  \right) /d\xi
 \Big] = \phi(\xi)$,  provided that the probability density exists for the probability on left hand side. 
  For use of this kind of notation, see for instance remarks in Appendix B of \cite{Touchette}. 

The authors in \cite{Sethuraman2013}  found an expression of the rate function $\phi(\xi)$ in terms of that 
for the process empirical density, which we recall here. Let 
$M_1=M_1(\rho_-,\rho_+)$ denote the space of functions which equals $\rho_-$ (resp. $\rho_+$) for all 
$x\leq x_*$ (resp. $x \geq x^*$) for some $x_*\leq x^*$, $D(M_1;[0,T])$ the set of right continuous 
trajectories on $M_1$ with left limits, and $C_K^{2,1}(\mathbb{R}\times [0,T])$ 
the space of compactly supported functions twice and once
continuously differentiable
in $x$ and $t$.\footnote{Note $x$ here represents the scaled and continuous 
 variable whereas up to here $x$ refers to a discrete site.}
 For $\mu\in D(M_1,[0,T])$,  let us define the linear functional on $C_K^{2,1}(\mathbb{R}\times [0,T])$, 
 \begin{align}
  l(\mu;G) 
  = 
  \int G(x,T)\mu_T(x) dx - \int G(x,0)\mu_0(x) dx 
  -
  \int_0^T\int \mu_t(x)\left(\frac{\partial}{\partial t}+\frac12\frac{\partial^2}{\partial x^2}\right)G(x,t) dx dt,
 \end{align} 
where $\mu_t(x)=\mu(x,t)$ and $G\in C_K^{2,1}(\mathbb{R}\times [0,T])$. 
Besides,  the process empirical density is defined by 
\begin{equation}
 \mu^N(s,x,\eta) = \sum_{k\in\mathbb{Z}} \eta_{k}(N^2s)1_{[k/N,(k+1)/N)}(x)
\end{equation}
where $x\in\mathbb{R},s\in [0,T), 0<T<\infty$ and $N$ is a large integer tending to infinity. 
The results of \cite{KOV,Sethuraman2013} (Corollary 1.4 in \cite{Sethuraman2013}) say that for the process 
starting from a local  equilibrium measure, 
corresponding to a density profile $\gamma\in M_1$, the large deviation principle holds and that 
the rate function $I(\mu)$ for the process empirical density is given by 
\begin{equation}
\label{Imu}
 I(\mu) = I_0(\mu) + h(\mu_0;\gamma) 
 \end{equation}
 where 
 \begin{align}
 I_0(\mu) 
 &= 
 \sup_{G\in C_K^{2,1}(\mathbb{R}\times [0,T])}
           \left\{l(\mu;G)-\frac12\int_0^T\int\mu_t(1-\mu_t) G_x^2(x,t) dxdt\right\}, \\
 h(\mu_0;\gamma) 
 &=
 \sup_{\phi_0,\phi_1\in C_K(\mathbb{R}) }
          \left\{\int \mu_0(x)\phi_0(x)dx + \int(1-\mu_0(x))\phi_1(x)dx  \right. \\
         &\quad \left. -\int\log[ \gamma(x)e^{\phi_0(x)}+(1-\gamma(x))e^{\phi_1(x)}]\right\}. 
 \end{align}
 Here $C_K(\mathbb{R})$ is the space of compactly supported continuous functions. 
 Then  it was shown in \cite{Sethuraman2013} (Theorem 1.5) that the rate function $\phi(\xi)$ for the 
 tagged particle is given in terms of $I(\mu)$  by
 \begin{equation}
  \phi(\xi) = \inf_{\mu\in D(M_1,[0,T])}\left\{I(\mu): \int_0^{\infty} (\mu_T(x)-\mu_0(x)) dx = \int_0^{-\xi} \mu_T(x) dx \right\}.
  \label{phiI}
 \end{equation}  

By using the representation above, the authors of \cite{Sethuraman2013}  have given some bounds on the rate function, 
studied its asymptotic growth and its behavior  near its vanishing minimum  $\phi(\xi_0) = 0$, which is attained when 
$\mu_t(x)$ becomes $\rho(x,t)$ in (\ref{rhoxi0}), i.e., at $\xi_0$ given by (\ref{AveragePosition}). 
  
  \vskip 0.3cm
  
In this article we shall  calculate explicitly the rate function $\phi(\xi)$ when the system is prepared in 
a step initial profile with densities $\rho_-$ and $\rho_+$ on the left and on the right of the origin where
the tracer is initially located. In fact we will find a formula for the distribution of the tagged particle for 
finite $t$ and $x$. Then,  the rate function $\phi(\xi)$ will  be extracted
rather straightforwardly from it. 

 Knowing the large deviation function  $\phi$  will allow us  to compute
 the cumulants of the tracer's position $X_t$ in the limit  $t \to \infty$. 
These results have been announced
in the letter \cite{IMSprl} and the  purpose of the present work is to
provide the reader with proofs of our claims and some generalizations of our results.
Besides the derivations of our results have been simplified considerably and we present  here these improved arguments.  


\subsection{Current and height variables}
\subsubsection{Definition}
To study properties of the tagged particle position $X_t$, it is  useful to consider  a 
related quantity, $Q(x,t)$, which is the  time-integrated current
 that has flown through the bond $(x,x+1)$ between time 0 and $t$, i.e.  $Q(x,t)$
  is equal to the total number of particles that have jumped
 from $x+1$ to $x$  {\it minus} the total number of particles that have jumped
 from $x$  to $x+1$ during the time interval $(0,t)$.

 Let us define also the local  height function
  $N(x,t)$  over the site $x$ at time $t$ by 
\begin{equation}
 N(x,t) = N_t+ \begin{cases}
                             + \sum_{y=1}^x \eta_y(t) \,, & x>0, \\
                              0, & x=0, \\
                             -  \sum_{y=x+1}^0 \eta_y(t) \,, & x<0 ,       
                        \end{cases}
\label{def:N}
\end{equation}
where $N_t=Q(0,t)$ is another notation for the time-integrated current through the bond
 $(0,1)$ during the time interval $[0,t]$. 
   
\vskip 0.3cm
\noindent
 {\it Remark.}  In  the usual   mapping between the exclusion
 process in one-dimension and a fluctuating interface in a  solid-on-solid
 model,  it is the  function $h(x,t) = N(x,t)-  x/2$ that  is 
 defined  as the local height of the interface.
 For the case of asymmetric hoppings, 
 in a scaling limit,  the  interface  $h(x,t)$ satisfies
 the Kardar-Parisi-Zhang equation \cite{BG1997}.
\vskip 0.3cm


By  particle conservation, $Q(x,t)$ and  $N(x,t)$ are related as follows:
At  $x = 0$, $Q(0,t) = N(0,t) = N_t$.  
For $x > 0$,
$Q(x,t) = Q(0,t) +  \sum_{y=1}^x \big(\eta_y(t) -\eta_y(0)\big)$
 or equivalently  
\begin{equation}
 N(x,t) = Q(x,t) +  \sum_{y=1}^x \eta_y(0) \,.
 \label{NversusQ1}
\end{equation}
And for $x < 0$, 
\begin{equation}
 N(x,t) = Q(x,t) -  \sum_{y=x+1}^0 \eta_y(0) \, .
 \label{NversusQ2}
\end{equation}


In this article we focus on the {\it two-sided Bernoulli initial condition}: 
initially all sites are independent,  a site with  $x<0$ is occupied with 
probability $\rho_-$ and a site  $x\geq 0$ is occupied with probability 
$\rho_+$.   

For this initial condition, the current $Q(x,t)$ and the local height $N(x,t)$ satisfy 
some simple properties  under    symmetries  that  leave  the SEP
 dynamics invariant  \cite{Doucot}.  In particular,
  spatial parity  (i.e. left-right
 symmetry)  implies 
 \begin{equation}
    N(-x,t, \rho_+, \rho_-) \overset{d}{=} -  N(x,t, \rho_-, \rho_+)
   \label{SpaceParity}
 \end{equation}
  where $\overset{d}{=}$ means the equality in distribution and, for clarity sake,
 we have written explicitly the dependence on boundary densities.
 Similarly, from particle-hole conjugation, we have  
   $Q(x,t, \rho_+, \rho_-) \overset{d}{=} - Q(x,t, 1- \rho_+,1- \rho_-)$,  and therefore
 \begin{equation}
    N(x,t, \rho_+, \rho_-) \overset{d}{=}  x - N(x,t,1- \rho_+,1- \rho_-) .
 \end{equation}

\subsubsection{Relation between the local height and the tracer's position}

 At time $t=0$, we   tag the  particle which is 
  closest to the origin from the right and call it the
 'tracer'.  Its initial position is 
 $X_0 \ge 0$ and its  position at time $t$ is denoted 
 by  $X_t$. Using  particle number conservation, 
 the  probability distribution
 of the  position  $X_t$ of a tagged particle  is related to 
 the distribution 
 of  the local   height function  $N(x,t)$  by the following identity 

\begin{equation}
 {\quad  \mathbb{P}[X_t \leq x] = \mathbb{P} [N(x,t)  >  0]  } \, . 
\label{XversusN}
 \end{equation} 

 This formula  will allow us to derive the statistical properties
 of  $X_t$  from  those of  $N(x,t)$, which are more tractable because
 the  height functions  are local observables  contrarily to 
 $X_t$, which is drifting away with time.
  We recall how one finds Eq.~(\ref{XversusN})
 for completeness sake  \cite{IMSprl}.
  Consider a site  $x> X_0$, located
 to the right of  the initial
 position  of the tracer.
 For  the tracer $X_t$ to  be strictly
 to the right of $x$  at time $t$,
 it is necessary that all the particles that were initially  between  $X_0$  
 and $x$ have crossed the bond  $(x,x+1)$
 from left to right.
 This means that the  total current $Q(x,t)$ at $x$ has to be less than
 $- \sum_{i=X_0}^x \eta_i(0) = - \sum_{i=1}^x \eta_i(0)$ (using the
 fact   that  all sites between 1 and $X_0 -1$ are empty at $t =0$).
 Hence,  for   $x \ge  X_0$, we have 
\begin{eqnarray}
 \hbox{Prob}\left( X_t > x \right) &=& 
 \hbox{Prob}\left(Q(x,t) \le - \sum_{i=1}^x \eta_i(0) \right)  \nonumber  \\
 &=&  \hbox{Prob}\left(N(x,t) \le 0 \right) \, . 
\label{eq:dte}
\end{eqnarray}
This equation is equivalent to (\ref{XversusN}).

\vskip 0.2cm
     More generally, 
 at $t = 0$, once  the particle closest to
 the origin in the region  $x \ge 0$ is selected as the tracer, 
 all the particles in the system  can be  labeled as~: 
$$\ldots < X_2 < X_1 <X_0 <X_{-1}<X_{-2} < \ldots $$
Then, the position  $X_m(t)$  of the $m$-th tagged particle at time $t$ 
 is related to the local height variables as follows~:
\begin{align}
\mathbb{P}[X_m(t) \leq x] 
&= 
\mathbb{P} [N(x,t)  \geq m] \, .  \notag\\
\label{Mth-Tracer}
\end{align}

\subsubsection{Large deviation for the current and the height}
In \cite{Sethuraman2013}, the authors have shown that 
 in the long-time limit, the current $Q(x,t)$ satisfies the large deviation principle
 \begin{equation}
 \mathbb{P}\left( \frac{Q(x,t)}{\sqrt{t}} = q  \right)
  \simeq  \exp[-\sqrt{t} \Psi(\xi,q)]   \,  \,  \,  \, 
 \hbox{ with }  \,  \,  \,  \, 
  \xi = - \frac{x}{\sqrt{4 t}}, 
\label{def1:PHI}
\end{equation}
where $\Psi(\xi,q)$ is the rate function of  $Q(x,t)$. 
They also gave a formula for the large deviation function for the current in terms of the 
rate function for the process empirical measure $I(\mu)$ in (\ref{Imu}) to be 
  \begin{equation}
  \Psi(\xi,q) = \inf\left\{I(\mu): \int_{-\xi}^{\infty} (\mu_T(x)-\mu_0(x)) dx =  q \right\}.
 \end{equation}  
 The above statement  was made only for $x=0$ case
in \cite{Sethuraman2013}. It also holds  for general $x$  by translational invariance.

By applying the same arguments,  
one can show that the height $N(x,t)$ also satisfies the large deviation principle
\begin{equation}
 \mathbb{P}\left( \frac{N(x,t)}{\sqrt{t}} = q  \right)
  \simeq  \exp[-\sqrt{t} \Phi(\xi,q)]   \,  \,  \,  \, 
 \hbox{ with }  \,  \,  \,  \, 
  \xi = - \frac{x}{\sqrt{4 t}} 
\label{def:PHI}
\end{equation}
where $\Phi(\xi,q)$ is the rate function of the local height variable $N(x,t)$. 
In addition, in terms of $I(\mu)$ this can be written as 
\begin{equation}
  \Phi(\xi,q) = \inf\left\{I(\mu): \int_{-\xi}^{\infty} (\mu_T(x)-\mu_0(x)) dx + \int_0^{-\xi} \mu_0(x) dx=  q \right\}.
\label{PhiI}
\end{equation}

\vskip 0.2cm
Let us denote the expectation value by the bracket $\langle \cdots \rangle$. 
 The  characteristic function of $N(x,t)$  is defined as  $\langle e^{\lambda N(x,t)} \rangle$ for 
 all  $\l \in \mathbb{R}$,  -- indeed,  by  using  \eqref {Mth-Tracer} and ignoring the
 exclusion condition, we  note that  the tails  of the distribution of $N(x,t)$ are bounded above
 by a Gaussian.  (Moreover we will find an  explicit
 expression for this  function  which admits an analytic continuation 
 to $\l\in\mathbb{C}$.)  By taking derivatives with respect to  $\lambda$ (assuming  smoothness),  
 the characteristic function generates  all the moments
 of the random variable  $N(x,t)$. By Varadhan's lemma,  see for instance Section 4.3 in \cite{DemboZeitouni},
 the large deviation principle \eqref{def:PHI}
 implies  that the characteristic function behaves as 
\begin{equation}
  \langle e^{\lambda N(x,t)} \rangle \simeq  e^{- \sqrt{t} \mu(\xi,\lambda)} 
 \quad \hbox{ for}  ~ \l\in\mathbb{R} ~{\rm and} ~t \to \infty \,. 
\label{def:mu}
\end{equation}
 The asymptotic cumulant generating function $\mu(\xi,\lambda)$ and the 
  rate function  $\Phi(\xi,q)$  are related
 by a  Fenchel-Legendre transform~:
\begin{equation}
   \Phi(\xi,q)  = \max_{\lambda}\left(\mu(\xi,\lambda) + 
  \lambda q  \right) . 
\label{PhiVERSUSmu}
\end{equation}

 \subsection{Exact expressions of the generating function and cumulants}
 \subsubsection{Finite time formula} 
 The central result of this work is an  exact expression for the  characteristic function of $N(x,t)$ 
valid for any  finite values of $x$ and $t$, that can be expressed
in terms of a Fredholm determinant.

\begin{theorem}\label{thm:GF}
For the two-sided Bernoulli initial condition with density $\rho_-$ (resp. $\rho_+$) to the left (resp. right), 
the characteristic function of $N(x,t), x\in\mathbb{Z}, t\geq 0$, defined in (\ref{def:N}) can be written as 
\begin{equation}
\langle e^{\l N(x,t)} \rangle 
=  
\det(1+\o K_{x,t})_{L^2(C_0)} \cdot M_0(\l), 
\label{GFfinitetime}
\end{equation}
where the determinant on the right hand side is a Fredholm determinant with the kernel, 
\begin{equation}
  K_{x,t}(\xi_1,\xi_2) = \frac{\xi_1^{|x|} e^{\ve(\xi_1) t}}{\xi_1\xi_2+1-2\xi_2},
  \label{Kxt}
\end{equation} 
\begin{equation}
\o = \rho_+(e^{\l}-1)+\rho_-(e^{-\l}-1)+\rho_+\rho_-(e^{\l}-1)(e^{-\l}-1), 
\label{omega}
\end{equation} 
and 
\begin{eqnarray}
  M_0(\l) &=& \begin{cases}
         \left(1+\rho_+(e^{\l}-1)\right)^x  \quad\quad   {\rm  for }   \quad& x\geq 0, \\
         \left(1+\rho_{-}(e^{-\l}-1)\right)^{-x}   \quad  \rm{ for  }  \quad&   x < 0 .   
                        \end{cases}
\end{eqnarray} 
Here $\ve(\xi) = \xi+1/\xi-2$ and 
$C_0$ is a contour around the origin with the radius small enough so that it does 
not include the poles from the denominator of the kernel (\ref{Kxt}).
\end{theorem}

\noindent
{\it Remark.}  The characteristic function $\langle e^{\l N(x,t)} \rangle$ on the left hand side 
is originally defined for $\l\in\mathbb{R}$. But the kernel  $K_{x,t}$ being smooth in the vicinity 
of  the  small contour $C_0$, the corresponding operator is trace-class (\cite{Lax}, page 345)
and therefore the Fredholm determinant  \eqref{GFfinitetime} on the right hand side 
is well defined for all values  of $\omega$ \cite{Bornemann2010, TW2008b}, or equivalently for 
all values of $0\leq \rho_{\pm} \leq 1, \l\in\mathbb{C}$. 
  

\medskip
The key for the proof of the above theorem is the following formula for the moments of $N(x,t)$. 
\begin{proposition} 
\label{prop:Nmom}
For $x\geq 0, t\geq 0$, the $n$-th moment of $N(x,t)$ is given by 
\begin{equation}
 \langle N(x,t)^n \rangle 
 =
 \sum_{k=0}^n m_{n,k} J_k(x,t),
 \label{NmomJ}
\end{equation} 
where $m_{n,k}$ is defined through 
\begin{equation}
 \sum_{n=0}^{\infty} \frac{\l^n}{n!}m_{n,k} = \frac{\o^k}{k!}(1+\rho_+(e^{\l}-1))^x
 \label{Mk}
\end{equation}
and $J_k(x,t)$ is given by 
\begin{equation}
\label{Jk}
J_k = J_k(x,t)
 = 
 \int_{C_0} \cdots \int_{C_0} 
 \prod_{i<j} \frac{\xi_i-\xi_j}{1+\xi_i \xi_j -2\xi_j} 
 \prod_{i=1}^k \frac{\xi_i^x e^{\ve(\xi_i) t}}{(1-\xi_i)^2} d\xi_i .
\end{equation}
Here $\ve(\xi)$ and $C_0$ are the same as in Theorem \ref{thm:GF}. 
The expression for $x<0$ case is found by applying the symmetry (\ref{SpaceParity}). 
\end{proposition} 

We will prove this proposition by first considering a $\tau$-deformed moment for the ASEP in Section 
\ref{sec:momASEP} and then taking the symmetric limit in Section \ref{sec:symlim}. 
To prove Theorem \ref{thm:GF} we also need 

\begin{proposition}
\label{prop:JFdet}
The generating function of  the integrals $\{J_n\}_{n\in\mathbb{N}}$ can be expressed as a Fredholm  determinant
\begin{equation}
\label{Jgen}
 \sum_{n=0}^{\infty} \frac{\zeta^n}{n!} J_n 
 =
 \det(1+\zeta K_{x,t})_{L^2(C_0)}, 
\end{equation}
where $\zeta\in\mathbb{C}$ and  $K_{x,t}$ and $C_0$ are as explained in Theorem \ref{thm:GF}. 
\end{proposition}
\noindent
The proof of this proposition will be given in Section \ref{sec:JFdet}. 
Using Propositions \ref{prop:Nmom} and \ref{prop:JFdet},  
Theorem \ref{thm:GF} is readily proved.  

\medskip
\noindent
{\it Proof of Theorem \ref{thm:GF}}.
For $x\geq 0$ we have
\begin{align}
 \langle e^{\l N(x,t)} \rangle
 &= \sum_{n=0}^{\i} \frac{\l^n}{n!} \langle N^n \rangle
 =
 \sum_{n=0}^{\i} \frac{\l^n}{n!} \sum_{k=0}^n m_{n,k} J_k 
 =
 \sum_{k=0}^{\i} ( \sum_{n=0}^{\i} \frac{\l^n}{n!}m_{n,k} ) J_k \notag\\
 &=
 \sum_{k=0}^{\i} \frac{\o^k}{k!} M_0(\l) J_k 
 =
 \sum_{k=0}^{\i} \frac{\o^k}{k!}  J_k \cdot M_0(\l)
 =
 \det(1+\o K_{x,t})_{L^2(C_0)} \cdot M_0(\l). 
\end{align}
We used Proposition \ref{prop:Nmom} in the second equality and Proposition \ref{prop:JFdet}
in the last equality. This concludes the  proof of  (\ref{GFfinitetime}) for $x\geq 0$. The case $x<0$ is 
deduced by using left-right symmetry \eqref{SpaceParity}.
 \qed 
 

\medskip
For any finite values of  $t$ and $x$, the cumulants of $N(x,t)$ are obtained by  taking  the  logarithm 
of \eqref{GFfinitetime} and expanding the resulting formula with respect to $\lambda$. We have 
\begin{corollary}
For $x \geq 0$, we obtain 

\noindent
(a) 
\begin{equation}
\log \langle e^{\l N(x,t)} \rangle 
=
\sum_{n=1}^{\infty}\frac{(-1)^{n-1}\o^n}{n} I_n(x,t) + x\log(1+\rho_+(e^{\l}-1))
\label{logGF}
\end{equation} 
where $\o$ is given by (\ref{omega}) and 
\begin{equation}
   I_n  = I_n(x,t) =
 \Tr K_{x,t}^n
 = 
  \int_{C_0} \cdots \int_{C_0} 
  K_{x,t}(\xi_1,\xi_2) \cdots K_{x,t}(\xi_n,\xi_1) \prod_{i=1}^n d\xi_i . 
\label{def:In}
\end{equation} 
(b) 
The $n$-th cumulant of $N(x,t)$ is given by 
\begin{align}
\label{cumulantfinitetime}
 \langle N(x,t)^n \rangle_c  =
\sum_{l=1}^n  (-1)^l (l-1)! \left(\alpha_{n,l}(r_+,r_-) I_l(x,t)  + \alpha_{n,l}(1,0) x  \rho_+^l \right) 
\end{align} 
where 
\begin{align}
 \alpha_{n,l}(a,b)  
 = 
\sum_{\substack{\nu \vdash n \\ \nu=1^{l_1} 2^{l_2} \cdots \\ l_1+l_2+\cdots = l} }
 \frac{n!}{ \prod_{j=1}^n l_j!} \,  \prod_{j=1}^n 
\left(\frac{a + (-1)^jb }{j!}\right)^{l_j}.
\label{def:alphanl}
\end{align} 
(We recall that  symbol $\nu \vdash n$ means that $\nu = 1^{l_1} 2^{l_2} \cdots$
  is a partition of $n$, {\it i.e.}, $n = \sum_j j l_j$.)
  
\medskip  
\noindent 
The expressions for $x<0$ case can be found by applying the symmetry (\ref{SpaceParity}). 
\end{corollary} 

\medskip
\noindent
{\it Remark.} 
One can show the convergence of the series on the right hand side of (\ref{logGF}) for 
instance for $|\o|<\sqrt{2/\pi}$ by the arguments in \cite{DG2009}. But since the 
Fredholm determinant in (\ref{GFfinitetime}) is well defined for all $\o\in\mathbb{C}$, 
the exponential of the right hand side of (\ref{logGF}) can be analytically continued to 
all $\o\in\mathbb{C}$, see (3.3) in \cite{Bornemann2010}. 

\medskip
\noindent
{\it Proof.} (a) For the first term, on the right hand side we use 
\begin{align}
 \log (\det(1+\omega K_{x,t})) 
 &=
 \Tr \log(1+\omega K_{x,t})
 =
 \sum_{n=1}^{\infty} \frac{(-1)^{n-1} \omega^n}{n} \Tr K_{x,t}^n 
 = 
 \sum_{n=1}^{\infty} \frac{(-1)^{n-1} \omega^n}{n} I_n. 
 \label{kappagen}
\end{align}
The second term is trivial. 

\noindent
(b) This is basically due to the general relation between the moments and cumulants, 
which are recalled in Appendix \ref{sec:momcum}. The second term is exactly the 
consequence of Example 1 in the Appendix. The first term is closely related to 
Example 2 in the Appendix. First note that, if we introduce 
\begin{equation}
 r_{\pm} = \rho_{\pm}(1-\rho_{\mp}), 
\end{equation} 
$\o$ in (\ref{omega}) can be written as 
\begin{equation}
 \omega = r_+(e^{\l}-1) + r_-(e^{-\l}-1). 
\end{equation}
Since in (\ref{logGF}) $I_n$ appears with $\omega^n$ which is order $n$ in $r_{\pm}$, 
in (\ref{cumulantfinitetime}) $I_l$ should appear with $\alpha_{n,l}(r_+,r_-)$ which is of order $l$ in $r_{\pm}$. 
\qed

 \subsubsection{Long time limit} 
 We are interested in the long time behaviors on the scale where  $ \xi = - \frac{x}{\sqrt{4 t}}$ is kept constant. 
 Our result is the following. 
 
 \begin{theorem}
 \label{thm:cumgen}
 For the step Bernoulli initial condition with density $\rho_-$ (resp. $\rho_+$) to the left (resp. right), 
 the asymptotic cumulant generating function $\mu(\xi,\l)$ defined in (\ref{def:mu}) is given by  
\begin{align} 
\mu(\xi,\l) =
 \sum_{n=1}^{\infty}  \frac{ (-\omega)^n}{n^{3/2}}  A(\sqrt{n}\, \xi) \,
 +  \xi \log\frac{ 1+\rho_+(e^{\lambda}-1)} 
{1+\rho_-(e^{-\lambda}-1)}, \quad \xi\in\mathbb{R},
\label{cumgen}
\end{align}
with $\o$ given by (\ref{omega}) and 
\begin{equation} 
A(\xi)   
 = \frac{e^{-\xi^2}}{\sqrt{\pi}}+\xi(1-{\rm erfc}{(\xi)}) =
  \xi +  \int_{\xi}^\infty {\rm erfc}{(u)}  {\rm d}u . 
 \label{def:Axi}
\end{equation}
 \end{theorem} 
 
 \noindent 
 This  expression of $\mu(\xi,\lambda)$ can be recast in a more
 symmetric  form  using  
 \begin{equation}
  \Xi(\xi)  =  A(\xi) -  \xi\ = \int_{\xi}^\infty \erfc{(u)}  {\rm d}u =  \frac{1}{\sqrt{\pi}}\int_0^1 e^{-\xi^2/u^2}du
    \label{def:Xi} 
 \end{equation}
and  
\begin{equation}
  \Xi_n(\xi)  =  \frac{1}{\sqrt{n}}  \Xi(\sqrt{n}\xi).
    \label{def:Xin} 
 \end{equation}
 Note that $A(\xi)$ is an even function in $\xi$ and $\Xi(\xi)$ satisfies a relation 
 \begin{equation}
  \Xi(\xi)-\Xi(-\xi) = -2\xi. 
  \label{Xisym}
 \end{equation} 
 We have 
 \begin{equation}
  A(\xi) = \frac12(\Xi(\xi)+\Xi(-\xi)) = \Xi(-\xi) + \xi. 
 \end{equation}
 Then, \eqref{cumgen} can be rewritten as  
\begin{align} 
\mu(\xi,\l) 
 &=
 \sum_{n=1}^{\infty}  \frac{ (-\omega)^n}{ { 2} n}
  \left(\Xi_n(\xi) + \Xi_n(-\xi)\right) \,
 +  \xi \log\frac{ 1+\rho_+(e^{\lambda}-1)} 
{1+\rho_-(e^{-\lambda}-1)} \notag\\
&=
 \sum_{n=1}^{\infty}  \frac{ (-\omega)^n}{n}
  \Xi_n(-\xi) \,
 + 2 \xi \log (1+\rho_+(e^{\lambda}-1)) ~. 
\label{cumgenbis}
\end{align}
In the last equality we used $1+\omega = (1+\rho_+(e^{\l}-1))(1+\rho_-(e^{-\l}-1))$. 

The proof of Theorem \ref{thm:cumgen} is not difficult once we understand the asymptotics of 
 $I_n(x,t)$ in (\ref{def:In}) on the same scale. 
 \begin{proposition}
 \label{prop:Iasym}
For $x\ge 0$,
\begin{equation}
\label{Iasym}
 I_n(x,t) \sim 
  \sqrt{t} \Xi_n(-\xi)
\end{equation}
where $x=-2\xi\sqrt{t}$ (and thus $\xi < 0$) and $\Xi_n(\xi)$ is given by (\ref{def:Xin}). 
The symbol $\sim$ means that the ratio of the left and right hand sides go to unity as $t\to\infty$. 
 \end{proposition} 

\noindent 
The proof of this proposition will be given in Appendix \ref{sec:asym}. 

\medskip
\noindent
{\it Proof of Theorem \ref{thm:cumgen}.}
Inserting the asymptotic behavior (\ref{Iasym}) of $I_n$ in (\ref{logGF}), we find 
(\ref{cumgenbis}), which is equivalent to (\ref{cumgen}) for $x\geq 0$. 
 The case $\xi >0$ i.e.  $x<0$ results again from left-right symmetry \eqref{SpaceParity}. 
This completes the proof of Theorem \ref{thm:cumgen} 
\qed

\medskip
The rate  function $\Phi(\xi,q)$ of  the local height $N(x,t)$, is 
the Fenchel-Legendre Transform  of $\mu(\xi,\lambda)$ \eqref{PhiVERSUSmu}.
Note that our arguments to arrive at the formula for $\mu(\xi,\l)$ are 
purely based on the exact formula in Proposition \ref{thm:GF} for the generating function 
$\langle e^{\l N(x,t)}\rangle$ and its asymptotic analysis, not depending
 on the large deviation principle for $N(x,t)$. Hence one can establish the large deviation 
 principle for $N(x,t)$ by using the G\"artner-Ellis theorem, see for instance Section 2.3 in \cite{DemboZeitouni}. 
  
The long time asymptotics of the $n$-th cumulant can  also be found from (\ref{cumulantfinitetime}). 
\begin{corollary}
\begin{align}
\label{cumulant}
 \lim_{t\to\infty}\frac{\langle N(x,t)^n \rangle_c}{\sqrt{t}} 
 =
\sum_{l=1}^n  (-1)^l (l-1)! \left(\alpha_{n,l}(r_+,r_-)
 \Xi_l(-\xi)  -2\alpha_{n,l}(1,0) \xi \rho_+^l \right) 
\end{align} 
the coefficients  $\alpha_{n,l}(a,b)$  being defined in \eqref{def:alphanl}. 
\end{corollary} 
\noindent
The expression (\ref{cumulant}) corresponds to the expression (\ref{cumgenbis}) of the generating 
function. One could find a few other equivalent expressions using (\ref{Xisym}).

\subsection{The distribution function of the tracer and some physical consequences}
\subsubsection{Finite time formula} 
As a corollary of Theorem \ref{thm:GF}, we obtain
\begin{corollary}
For the step Bernoulli initial condition with density $\rho_-$ (resp. $\rho_+$) to the left (resp. right), 
the distribution function of the tagged particle is written as 
\begin{equation}
 \mathbb{P}[X_t\leq x]
 =
 \int_{C_0} \frac{dz}{1-z} \det(1+\omega K_{x,t})_{L_2(C_0)}M_0(\l) |_{e^{-\l}=z}
 \label{distXt}
\end{equation}
where $C_0, \omega, K_{x,t}, M_0(\l)$ are the same as in Theorem \ref{thm:GF}. 
\end{corollary}

\noindent
{\it Proof.} 
Setting $z=e^{-\l}$, the generating function $\langle e^{\l N(x,t)}\rangle$ becomes 
$\langle z^{-N(x,t)} \rangle= \sum_{n\in\mathbb{Z}} \mathbb{P}[N(x,t)=n] z^{-n}$,
where the convergence  of the series is ensured by comparing SEP with independent 
random walkers. 
By the Cauchy theorem, we find 
\begin{equation}
 \mathbb{P}[N(x,t)=n]
 =
  \int_{C_0} dz 
  \langle z^{N(x,t)} \rangle z^{n-1}
 =
 \int_{C_0} dz  z^{n-1}
   \det(1+\omega K_{x,t})_{L_2(C_0)} M_0(\l) |_{e^{-\l}=z}. 
\end{equation}
Due to (\ref{XversusN}), the distribution function of the tracer can be written as 
\begin{equation}
\mathbb{P}[X_t\leq x] =  \sum_{n=1}^{\infty}\mathbb{P}[N(x,t)=n].
\end{equation}
Combining the above two, we find (\ref{distXt}). 
\qed

\noindent
{\it Remark.} Similar formulas have been found for ASEP \cite{TW2008b, TW2010a, BCS2014}.

\subsubsection{Large deviations} 
 Using the expressions  (\ref{phiI}) and (\ref{PhiI}) of the large deviations of $X_t$ and $N(x,t)$ 
in terms of $I(\mu)$ one can establish the simple relation between 
the rate function $\phi(\xi)$   of $X_t$, defined in  \eqref{def:phi} and  $\Phi(\xi,q)$, 
\begin{equation}
{ \quad \phi(\xi) =   \Phi(\xi,q =0) } ~. 
\label{PetitphiGrandPHI}
 \end{equation}
The relation can be more heuristically derived thanks to the identity  (\ref{XversusN}).  
Indeed, if  $x <  x_0$, the relation  \eqref{XversusN}
 can be written as, at the  large deviation level (keeping only exponentially
 dominant contributions), 
 $$ \int_{-\infty}^\xi   \exp[-\sqrt{t} \phi(u)]\,  {\rm d}u = 
 \int_0^{\infty} \exp[-\sqrt{t} \Phi(\xi, q)] \,  {\rm d}q .$$
The large deviation function $\phi(\xi)$  decreases from ${-\infty}$
 to $\xi_0$ (where it  vanishes).  Similarly $ \Phi(\xi, q)$
increases  when $q$  varies from 0 to ${\infty}$. 
 When $t \to \infty$, the
  dominant contributions to the integrals 
 come  from the boundaries, leading to $\phi(\xi) = \Phi(\xi, 0)$.
 For  $x >  x_0$, the reasoning goes along the same lines, starting from 
   $\hbox{Prob}\left( X_t >  x \right) 
 =  \hbox{Prob}\left(N(x,t) \le   0 \right).$ 

\vskip 0.2cm
\noindent
More generally if we consider a  particle  having  a label $m$
 scaling as $m = q \sqrt{t}$,  we can show along similar lines, starting from 
 Eq.~(\ref{Mth-Tracer}),  that the  large deviation
 function $\phi_m(\xi)$ of $X_m(t)$ is identical to  $\Phi(\xi,q)$. 

For the two-sided Bernoulli initial condition, the value of $\xi_0$ in (\ref{AveragePosition}) 
 can  be determined easily as follows. The average  density profile $ \rho(x,t)$ 
satisfies the diffusion equation and is   given by
 $$ \rho(x,t) =  \rho_{-} + \frac{\rho_{+} - \rho_{-}}{2} \hbox{erfc}(\xi).$$
By the particle number conservation, we find that $\xi_0$ is the unique solution of 
\begin{equation}
  2\xi_0  \rho_{-}  = 
 (\rho_{+}  - \rho_{-})  \int_{\xi_0}^\infty \erfc{(u)}  {\rm d}u
\label{value:xi0}
\end{equation}
where the complementary error function is defined by 
\begin{equation}
\erfc{(z)} = \frac{2}{\sqrt{\pi}} \int_z^\infty e^{-u^2}  {\rm d}u \, .
\label{def:erfc}
\end{equation}
 If the system is initially at equilibrium with uniform density
 $\rho$, then  $\xi_0$ vanishes. Otherwise, the tracer is  
  subject to the thrust of the hydrodynamic flow towards
  the low density region, 
 and its  average position  drifts away from the
 origin, growing as $t^{1/2}$.

\vskip 0.2cm

 Equations \eqref {PhiVERSUSmu} and \eqref{PetitphiGrandPHI}  
 provide us with an implicit representation of  the rate function $\phi(\xi)$. 
By the Varadhan lemma, 
the  large deviation principle (\ref{def:phi}) implies that the characteristic  function of 
$X_t$ behaves, in the long time limit, as 
\begin{align}
\langle e^{s X_t} \rangle \simeq  e^{- \sqrt{t} C(s)}
  \quad {\rm for } ~  s \in\mathbb{R} ~{\rm and} ~t \to \infty ,
\end{align}
and that  $C(s)$ is given by the Fenchel-Legendre transform of $\phi(\xi)$, i.e., 
 \begin{align}
  C(s) = \inf_{\xi} \left( 2s \xi  + \phi(\xi)  \right).
  \label{Cs}
\end{align}

In this asymptotic regime, the knowledge of the cumulant generating function
 $C(s)$ yields, by expanding around $s =0$ (assuming  smoothness),  
 all the cumulants of the tracer's position $X_t$ 
when $ t \to \infty$ \cite{IMSprl}. For example, in  the  equilibrium  case 
 $\rho_{-} =  \rho_{+} = \rho$, we obtain 
\begin{eqnarray}
  \frac{\langle X_t^2 \rangle_c}{\sqrt{4 t}}  &\to& 
   \frac{1-\rho}{\rho \sqrt{\pi} } , \label{2nd}\\  
 \frac{\langle X_t^4 \rangle_c}{\sqrt{4 t}}    &\to& 
  \frac{1-\rho}{\sqrt{\pi} \rho^3}
   [1-(4-(8-3\sqrt{2}) \rho)
  (1-  \rho)+\dfrac{12}{\pi}(1-\rho)^2] \label{4th}, 
\end{eqnarray}
when $t\to\infty$. The second cumulant (\ref{2nd}) was proved already in \cite{Arratia1983}
while the fourth (\ref{4th}) was first found by a perturbative calculation using the macroscopic 
fluctuation theory in \cite{KMSprl} and was confirmed in our work. 
 Higher cumulants and non-equilibrium formulas for 
 the variance are given in \cite{IMSprl}.

 Furthermore, we proved in  \cite{IMSprl} that the rate function 
 $\phi$ satisfies a symmetry relation, 
reminiscent of  the  fluctuation theorem  \cite{Gallavotti,LebSpohn}
\begin{equation}
 \phi(\xi)-\phi(-\xi)= 2 \xi \log\frac{ 1 - \rho_{+}}{1 - \rho_{-}}.
\label{FT}
\end{equation}
 In contrast to the usual fluctuation theorem, this formula does not involve
  a time reversal of  an external drive, but rather 
 a mirror image of  the initial condition.
  We remark that this  identity implies  the Einstein
 fluctuation-dissipation relation  \cite{Ferrari,Olla}


\section{Deformed moment formula for ASEP}
\label{sec:momASEP}
   
   To find a formula for the moment of $N(x,t)$, a natural strategy is to write down the evolution 
   equation for the moments of $N(x,t)$ and then solve it. 
   For the SEP $(\tau =1)$, the evolution equation for the $k$-th moment involves only moments and 
   correlations of  order $\le k$ but it becomes complicated when $k$ becomes large and seems difficult
   to solve directly. 
   But this difficulty can be circumvented by using a remarkable  property known as {\it duality}
   for asymmetric exclusion and then taking the symmetric limit. 

\subsection{Duality}

  For the symmetric exclusion process, the
   $n$-point correlation functions
  of the density {\it i.e.},  observables  of the type
  $$C(x_1, x_2, \ldots, x_n;t)   = \langle \eta_{x_1}(t)\eta_{x_2}(t)
 \ldots  \eta_{x_n}(t)\rangle 
  \quad  \hbox{ for }  \quad  x_1 < x_2 < \ldots < x_n \, , $$
  satisfy  the same evolution equations as the probability distribution
 of $n$ particles governed by  the SEP dynamics. This property
 is known as {\it self-duality} \cite{Liggett1985,Schuetz1997a}.

  Here,  we wish to calculate  correlations
  involving the variable $N(x,t)$, rather than   density
  correlations. Fortunately,  it can  be  shown that 
  for  the  ASEP,  when  $\tau <1$, 
   the  observable   $N(x,t)$,
   satisfies a striking  self-duality property
  \cite{Schuetz1997a,ImamuraDuality,BCS2014}.

 \begin{proposition} 
 For $x_1 < x_2 < \ldots < x_n$,   $n$-point correlations  of the type 
 \begin{equation}
 \phi(x_1,\ldots,  x_n; t)=  
  \langle \tau^{N(x_1,t)} \ldots  \tau^{N(x_n,t)} \rangle
\label{def:tau-corr}
\end{equation}
(which will be referred to as the $\tau$-correlations of order $n$ in the sequel) satisfy 
 the Kolmogorov forward equations for 
  the ASEP with a finite number $n$ of particles located at 
  $x_1,  \ldots,  x_n$.  
  \label{prop:ASEPduality}
  \end{proposition} 
  
This proposition  will be proved in the rest of this subsection.  
We emphasize  that the formula \eqref{def:tau-corr} makes sense only for 
$\tau \neq 1$ and can not be  applied to the symmetric 
case in a straightforward 
manner.  The duality for ASEP results from
 a fundamental quantum group symmetry of the Markov generator
 and has been shown  formally in various works
 \cite{Schuetz1997a,ImamuraDuality,Giardina1,Giardina2,Giardina3}.
Once found, this property can be verified directly without referring
  to this underlying symmetry. For instance, 
  a  direct proof of duality is  given 
 in \cite{BCS2014}, but it is restricted to the case 
  $\rho_+ =0$ or $\rho_-=0.$ 
 Here, we consider a system with  finite non-vanishing
  density of particles in both directions. Therefore, we shall 
 give an original proof of duality, using  stochastic calculus \cite{ItoMcKean,Karatzas,Klebaner}. 
 One could use a martingale introduced in \cite{Gartner1988} but our proof is more elementary. 

\vskip 0.2cm

\noindent 
{\it Proof of Proposition \ref{prop:ASEPduality}}. 
To appreciate the essential points of the proof, we first consider the $n=1,2$ cases. 
The one-point  $\tau$-correlation function is 
 $$  \phi(x;t) =  \langle \tau^{N(x,t)} \rangle  \, .$$
  Between $t$ and $t+dt$ , the variation of $  \phi$ is given by
 \begin{equation}
\phi(x;t+dt)  -  \phi(x;t) =
  \langle \tau^{N(x,t+dt)} - \tau^{N(x,t)}  \rangle 
 =  \langle \tau^{N(x,t)} \left(  \tau^{dN(x,t)} -1 \right)   \rangle 
\label{deltaphi}
\end{equation}
 Using  (\ref{NversusQ1}) and (\ref{NversusQ2}),  we observe that 
 between $t$ and $t+dt$, we have $ dN(x,t) = dQ(x,t) $ and therefore 
\begin{equation}
\tau^{dN(x,t)} -1 = \begin{cases}
                              \tau -1 \,, \quad \hbox{ with prob. }
\quad q\, \eta_{x+1}(t)(1 -  \eta_{x}(t))dt  \\
                           \frac{1}{\tau} -1  \,, \quad \hbox{ with prob. }
\quad  p\,\eta_{x}(t)(1 -  \eta_{x+1}(t))dt   \\
                           0      \,, \quad \hbox{ otherwise. } 
                        \end{cases}
\label{1-bodyevol}
\end{equation}
 Substituting in \eqref{deltaphi}, we obtain 
\begin{equation}
 \frac{ d \phi(x;t)}{dt} =  (p-q)
  \langle  \tau^{N(x,t)}  (\eta_{x+1}(t) -  \eta_{x}(t)) \rangle .
\end{equation}
Because the local occupation  is a binary variable  we can check
 that the following relation is identically true for any $x$ and $y$:
\begin{equation}
 (\tau -1) (\eta_{y}(t) -  \eta_{x}(t)) = \tau^{\eta_{y}(t)}
 + \tau\tau^{-\eta_{x}(t)} - (1 + \tau) \,. 
\label{binaryId}
\end{equation}
Inserting this relation (taking  $y = x+1$)
  in the previous equation and using the trivial identity $N(x+1,t)=N(x,t)+\eta_x(t)$, 
 we obtain 
\begin{equation}
 \frac{ d \phi(x;t)}{dt} =  q\phi(x+1;t) + p\, \phi(x-1;t) - 
  (p + q) \phi(x;t)
\end{equation}
which is identical to the evolution of a single particle  under ASEP dynamics.

 If we now consider a two-point  $\tau$-correlation function for
 $x_1 < x_2$, 
 $$  \phi(x_1,x_2;t) =  \langle \tau^{N(x_1,t)} \tau^{N(x_2,t)} \rangle  \, $$
 and write its time-evolution, we observe that if  $x_1$ and  $x_2$ 
 are not neighboring sites, the above calculations  remain valid and
 we obtain
\begin{eqnarray}
\frac{ d  \phi(x_1,x_2;t)}{dt} 
  = &&  q \phi(x_1+1,x_2;t) + p \phi(x_1-1,x_2;t) 
 + q \phi(x_1,x_2+1;t) +p \phi(x_1,x_2-1;t) \nonumber  \\
 &&  -  2 (p + q)  \phi(x_1,x_2;t)  \, , 
 \label{te_cor}
\end{eqnarray}  
 which is exactly  the  ASEP dynamics with  two particles at $x_1$ and $x_2$.
 The only special case is when $x_1 = x$ and $x_2 = x +1$. There, we check that
\begin{equation}
\tau^{dN(x,t)+ dN(x+1,t) } -1 = \begin{cases}
                              \tau -1 \,, \quad \hbox{ with rate }
\quad  q\,\eta_{x+2}(t)(1 -  \eta_{x+1}(t))+ q\,\eta_{x+1}(t)(1 -  \eta_{x}(t))  \\
                           \frac{1}{\tau} -1  \,, \quad \hbox{ with rate }
\quad  p\,\eta_{x+1}(t)(1 -  \eta_{x+2}(t))+
  p\,\eta_{x}(t)(1 -  \eta_{x+1}(t))   \\
                           0      \,, \quad \hbox{ otherwise. } 
                        \end{cases}
\nonumber
\end{equation}
 This leads to 
$$ \frac{d}{dt} \langle \tau^{N(x,t)} \tau^{N(x+1,t)}  \rangle 
  = (p -q) \langle \tau^{N(x,t)} \tau^{N(x+1,t)}
  ( \eta_{x+2}(t) -  \eta_{x}(t) )  \rangle  .$$
 By  using the relation  \eqref{binaryId} with $y = x+2$,
 the  r.h.s.  can be rewritten as 
 $q \phi(x,x+2;t) +p \phi(x-1,x+1;t) -  (p+q)  \phi(x,x+1;t)$,
  which is the ASEP dynamics for  two neighboring particles.

  We now consider the general 
 $n$-point $\tau$-correlation function  \eqref{def:tau-corr}. If
 none of  $x_1,x_2,\ldots,x_n$ are  neighboring points
 ({\it i.e.}, $x_{i+1} - x_{i} \ge 2$), then  
 each  factor $ \tau^{N(x_i,t)}$  will evolve as in \eqref{1-bodyevol}
 and  because the bond $(x_i, x_i +1)$ does  not interfere
  with the other  bonds, the total evolution of  $\phi(x_1,\ldots,  x_n; t)$
 is   obtained by adding individual contributions, leading to
\begin{align}
 & \frac{ d  \phi(x_1,\ldots,  x_n;t)}{dt} 
=   \nonumber \\
& q \sum_{i=1}^n \phi(x_1, \ldots, x_i+1, \ldots;t) + p
 \sum_{i=1}^n  \phi(x_1, \ldots, x_i-1, \ldots;t) 
  -  n (p+q)  \phi(x_1,\ldots,  x_n;t) .
   \nonumber
 \end{align}
   This equation represents the evolution of $n$ particles governed
 under  ASEP dynamics. One should now analyze the  cases when some
 of the $x_{i}$'s are 
 neighbors. Let us consider the extreme case of a single cluster
  $x_1=x, x_2 = x+1, \ldots,x_n = x+n-1$.
 Then, 
\begin{equation}
\tau^{dN(x,t)+ \ldots + dN(x+k -1,t) } -1 = \begin{cases}
                              \tau -1 \,, \quad \hbox{ with rate }
\quad  q \sum_{k=1}^n \eta_{x+k}(t)(1 -  \eta_{x+k-1}(t))  \\
                           \frac{1}{\tau} -1  \,, \quad \hbox{ with rate }
\quad  p \sum_{k=1}^n  \eta_{x+k-1}(t)(1 -  \eta_{x+k}(t))   \\
                           0      \,, \quad \hbox{ otherwise. } 
                        \end{cases}
\nonumber
\end{equation}
This leads to 
$$ \frac{d}{dt} \langle \tau^{N(x,t)} \tau^{N(x+1,t)} \ldots\tau^{N(x+n-1,t)}    \rangle 
  = (p -q) \langle \tau^{N(x,t)} \tau^{N(x+1,t)} \ldots\tau^{N(x+n-1,t)} 
  ( \eta_{x+n}(t) -  \eta_{x}(t) )  \rangle  .$$
 By  using the relation  \eqref{binaryId} with $y = x+n$,
 the term  $( \eta_{x+n}(t) -  \eta_{x}(t) )$ can  be re-exponentiated
 allowing us to rewrite the r.h.s. as 
 $q \phi(x,\ldots,x+n;t) +p \phi(x-1,\ldots,x+n-1;t) -  (p + q) 
  \phi(x,\ldots,x+n-1;t)$,
 which is  identical to  the ASEP dynamics  for
 a configuration of $n$  neighboring particles.
 To conclude,  one should also  examine the general case 
 with  $l$ clusters consisting
 of  $n_1,n_2,\ldots,n_l$ neighbors with  $n_1+n_2+\ldots+n_l = n$.
 Well separated clusters
 will evolve independently, and  a  cluster of $k$ neighbors
 located  between $x$ and $x+k -1$ 
 will produce  a term $(\eta_{x+k}(t) -  \eta_{x}(t) )$ in the
 evolution equation that can  be re-exponentiated  using
 \eqref{binaryId} with $y = x+k$. The result will again
 be identical to the  ASEP dynamics of  $n$ particles, located at 
  $x_1,  x_2 , \ldots, x_n$.
This ends the proof of the self-duality of ASEP. 
\qed

\medskip
Note that the evolution equations for the $\tau$-correlations are autonomous. We have 
\begin{proposition}
\label{prop:uniqueness}
The solution of the evolution equation in Proposition \ref{prop:ASEPduality} for (\ref{def:tau-corr}) is unique. 
\end{proposition}

\noindent 
{\it Proof. }
We can apply Proposition 4.9 in \cite{BCS2014}; the condition (30) of that proposition 
is satisfied because the number of particles between the origin and a site $x$ at $t=0$ 
is less than or equal to $|x|$.
\qed

\subsection{Contour integral formulas for the $\tau$-correlation function}
\label{tau-cor}

Using the fact that the  ASEP with $n$ particles
is  solvable  by Bethe Ansatz,  we  now  express
the   $\tau$-correlations  \eqref{def:tau-corr}   as a multiple contour
integral in the complex plane \cite{TW2008a,BC2014,BCS2014}.
The very  existence and validity
  of such  a formula   depends  crucially  on the initial
 conditions. We recall that we consider here step initial
 conditions with 
 initial densities  $\rho_{-}$ and  $\rho_{+}$  on the right and on the
 left of the origin.

\begin{proposition} \label{prop:ASEPcor}
We can write, for $x_1 < \cdots < x_n$, the following integral representation
\begin{equation}
\label{ASEPcor}
\langle \prod_{i=1}^n \tau^{N(x_i,t)} \rangle
= 
\tau^{n(n-1)/2} 
\prod_{i=1}^n
\left( 1-\frac{r_-}{\tau^i r_+} \right)
\int \cdots \int \prod_{i<j} \frac{z_i-z_j}{z_i-\tau z_j} \prod_{i=1}^n F_{x_i,t}(z_i) dz_i
\end{equation}
where 
\begin{equation}
 F_{x,t}(z) 
 = 
 \frac{\left( \frac{1+z}{1+z/\tau}\right)^x e^{\g(z) t}}
        {(1-\frac{z}{\tau\theta_+})(1-\frac{\theta_-}{z})z}
\end{equation}
with 
\begin{eqnarray}
 \th_{\pm} &=& \rho_\pm/(1-\rho_\pm)  \\
 \hbox{ and } \quad  \g(z) &=&  -\frac{q(1-\tau)^2 z}{(1+z)(\tau+z)} . \label{gamma}
\end{eqnarray}
The contour of each $z_i$ consists of two simple curves, $\Gamma_i^{(1)}$ and $\Gamma_i^{(2)}$, 
in the clockwise direction; $\Gamma_i^{(1)}$ contains  only the pole at
$-1$ and no other pole; $\Gamma_i^{(2)}$ encircles $\tau \th_+$, 
including $\{ \Gamma_j^{(2)}\}_{j>i}$, and no other pole, 
the integrations being performed in the order from $z_n$ to $z_1$.
See Figure \ref{fig:contours}.{\footnote{Throughout the paper all contour integrals are assumed to contain the factor $1/2\pi i$.}}

\begin{figure}[t]
\begin{center}
 \includegraphics[scale=1]{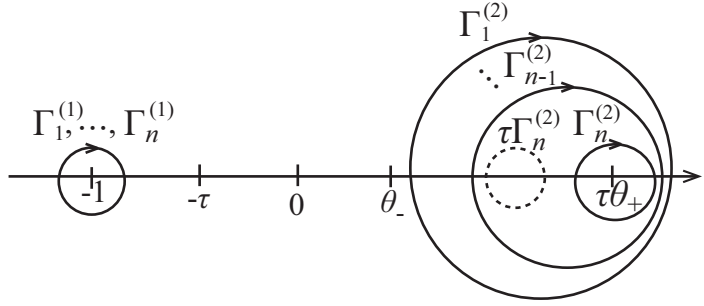}
 \end{center}
 \caption{Integration contours in the complex plane.  The contour of each $z_i$ consists of two curves $\Gamma_i^{(1)}$  and $\Gamma_i^{(2)}$: 
   the  $\Gamma_i^{(1)}$'s  encircle only the pole -1
   and no other pole, and are not nested. The $\Gamma_i^{(2)}$'s encircle $\tau\theta_+$ with
  a nesting condition  (see text) and exclude the poles at $-1,-\tau$ and  $\theta_-$.}
  \label{fig:contours}
 \end{figure}
\end{proposition}
\noindent
{\it Remark.} 
By the Cauchy theorem, one can switch to another set of contours s.t. the contour of $z_i$ includes  $-\tau, \theta_-$
and $\{z_j/\tau\}_{j<i}$  but not $-1, \tau\th_+$. 
The contours are now in the counter-clockwise direction and the integrations are performed in the order from $z_1$ to $z_n$. 
For the special case with $\theta_-=0$,  this nesting condition is equivalent to $z_i$ not including $\{\tau z_j\}_{j>i}$,
as employed in \cite{BCS2014}.  

\medskip
\noindent
{\it Proof.} 
The  integral  formula (\ref{ASEPcor})
(or equivalently,  \eqref{ASEPcor2})  can be proved by showing that
 the r.h.s. satisfies the Kolmogorov forward equation (or the master equation) for the $n$-particle ASEP 
 {\it together with} the  initial condition. The fact that (\ref{ASEPcor}) satisfies the master equation can be checked 
 by using the following relations, 
\begin{equation}
 \g(z) = p\frac{1+z/\tau}{1+z} + q\frac{1+z}{1+z/\tau} -(p+q), 
\end{equation}
\begin{equation}
 q\frac{(1+z_1)(1+z_2)}{(1+z_1/\tau)(1+z_2/\tau)}+p-\frac{1+z_2}{1+z_2/\tau}
 =
 \frac{(p-q)(z_1-\tau z_2)}{(1+z_1/\tau)(1+z_2/\tau)} .
\end{equation}
When particles are far apart, the first relation suffices. The second
 formula  is used to deal with the special case 
 when two particles are on neighboring sites.  For example, 
for  $n=2$, the factor $z_1-\tau z_2$ cancels the pole at $z_1=\tau z_2$;  the 
integrand becomes antisymmetric and the integral vanishes. 

  For verifying the initial condition, one first observes that when $t=0$ the essential singularities at 
  $z_i=-1,-\tau$ are absent and one can evaluate the contour integrals explicitly by taking
  either the pole at -1 or at  $-\tau$ depending on whether  $x_i$ 
  is positive or negative and taking the pole at around $\tau\th_+$ or $\th_-$.  
 For example, for  $n=1$ we get 
\begin{equation}
 \langle \tau^{N(x,0)} \rangle 
 =
 \begin{cases}
  (1-\rho_+ + \tau \rho_+)^x, & x\geq 0, \\
  (1-\rho_- +\frac{\rho_-}{\tau})^{-x}, & x\leq 0, 
 \end{cases}
\end{equation}
and for $n=2$, 
\begin{equation}
 \langle \tau^{N(x_1,0)+N(x_2,0)} \rangle 
  =
 \begin{cases}
   (1-\rho_+ + \tau^2 \rho_+)^{x_1} (1-\rho_+ + \tau \rho_+)^{x_2-x_1}, & 1\leq x_1 < x_2, \\
   (1-\rho_- +\frac{\rho_-}{\tau})^{-x_1} (1-\rho_+ + \tau \rho_+)^{x_2}, & x_1 \leq 0 < x_2, \\
   (1-\rho_- +\frac{\rho_-}{\tau})^{-x_1+x_2} (1-\rho_- +\frac{\rho_-}{\tau^2})^{-x_2}, & x_1<x_2<0.
 \end{cases}
 \end{equation}
For general $n$, for the case in which $x_1<\ldots <x_{l-1}\leq 0< x_l<\ldots x_n,1\leq l\leq n$, 
\begin{align}
 \langle \prod_{i=1}^n \tau^{N(x_i,0)} \rangle 
 &=
 \prod_{i=1}^{l-2} (1-\rho_-+\rho_-/\tau^i)^{x_{i+1}-x_i}(1-\rho_-+\rho_-/\tau^{l-1})^{-x_{l-1}} \notag\\
 &\quad\times(1-\rho_++\tau^{n-l+1})^{x_l}\prod_{i=l}^{n-1}(1-\rho_++\tau^{n-i}\rho_+)^{x_{i+1}-x_i}. 
\end{align}
They are the correct initial conditions because we are considering the Bernoulli measure 
with parameters $\rho_{\pm}$ to the right and left. 

Finally proposition \ref{prop:uniqueness} ensures that (\ref{ASEPcor}) is a representation of the 
$\tau$-correlation for ASEP with the two-sided Bernoulli initial condition.  
\qed 

We emphasize that the above integral formula for the $\tau$-correlation is valid for arbitrary $\tau <1$.  
When taking the symmetric limit, $\tau\to 1$, it is useful to rewrite it using  the change of variables, 
\begin{equation}
\label{cov}
 \xi_i = \frac{1+z_i}{1+z_i/\tau}. 
\end{equation}
Then the formula  (\ref{ASEPcor}) can be rewritten as 
\begin{align}
\langle \prod_{i=1}^n \tau^{N(x_i,t)} \rangle
&= 
(1-\tau)^n
\left( r_+-\frac{r_-}{\tau^i} \right) 
\tau^{n(n-1)/2} \int \cdots \int \prod_{1\leq i<j\leq n}
 \frac{\xi_i-\xi_j}{\tau-(1+\tau)\xi_j+\xi_i\xi_j} \notag\\
&\quad \times 
 \prod_{i=1}^n e^{\ve_{p,q}(\xi_i)t} \xi_i^{x_i} 
  \prod_{i=1}^n \frac{1}{(1-\xi_i-(1-\tau)\rho_+)(1-\xi_i+(1-1/\tau)\rho_-\xi_i)} d\xi_i
\label{ASEPcor2}
\end{align}
where $\ve_{p,q,}(\xi)=p\xi+q/\xi-(p+q)$ and the contours are obtained
from those  for the $z_i$ variables; they are around $0,1-(1-\tau)\rho_+$ with some 
nesting conditions. 
Note in the symmetric limit with $p=q=1$, $\ve_{p,q}$ tends to 
$\xi+1/\xi-2 = \ve(\xi)$ as defined in Theorem \ref{thm:GF}.

\subsection{Contour integral formula for the $n$-th moment}
The formula  (\ref{ASEPcor}) for the $\tau$-correlation 
has been obtained and proved to be valid only for  $x_1 < \cdots < x_n$.
In order to obtain an expression for the $\tau$-moment of order $n$ 
for $N(x,t)$  at a given point  $x$, we would need 
 to take  $x_i\equiv x$ in (\ref{ASEPcor}).
At this stage  of our discussions, there is absolutely no  guarantee that by
setting  $x_i\equiv x$ we would obtain a valid  and meaningful 
representation for the  $\tau$-moment of order $n$.
 But, in fact, it turns out that the resulting expression is the correct one. 

\begin{proposition}\label{prop:ASEPmoment}
For $n\in\mathbb{Z}$, the $n$-th $\tau$-moment of $N(x,t)$ is given by 
\begin{equation}
\langle \tau^{n N(x,t)} \rangle
= 
\prod_{i=1}^n\left( 1-\frac{r_-}{\tau^i r_+} \right) \tau^{n(n-1)/2} 
\int \cdots \int \prod_{i<j} \frac{z_i-z_j}{z_i-\tau z_j} \prod_{i=1}^n F_{x,t}(z_i) dz_i .
\label{ASEPmoment}
\end{equation}
The function $F_{x,t}(z)$ and the contours of $z_i$ are the same as in Proposition \ref{prop:ASEPcor}.
\end{proposition} 

This issue of finding $\tau$-moment formula from $\tau$-correlation was already noted 
in \cite{ImamuraDuality, BCS2014} where the case with 
 only a finite number of  particles on the left side  was treated.  
There another duality function was introduced and it was shown that the $\tau$-moment can be 
written as a sum of the correlation functions of this modified quantity evaluated at different points. 
However, the  arguments  developed in \cite{ImamuraDuality, BCS2014} are specific for
 the case with only a finite number of 
 particles to the left of the origin; these arguments   do
 not work for our case with finite densities  on the both sides. 

Here we give a different approach. We will write down the time evolution equation of the $\tau$-moment 
and show that the above formula satisfies that equation. To do so we need to write down the 
evolution equation for a more general case than the one covered by  duality, Proposition \ref{prop:ASEPduality}. 
Suppose that $x_j, j=1,\ldots,n$ are grouped into the form $(y_i,m_i), 1\leq  i\leq k$ with multiplicity $m_i$, i.e., 
$x_j=y_i$ for $\sum_{l=1}^{i-1} m_l < j\leq \sum_{l=1}^i m_l, 1\leq i\leq k$.  
\begin{proposition} 
\label{prop:eeASEPmoment}
The time evolution equation for the $n$-th moment is given by 
\begin{align}
& \quad \frac{d}{dt} \langle \tau^{n N(x,t)} \rangle =  \nonumber \\
& 
   \Big\langle \tau^{(n-2) N(x,t)} 
   \left(a_n \tau^{2 N(x,t)}   + b_n  \tau^{N(x,t)+ N(x-1,t)}  
   + c_n   \tau^{N(x,t)+ N(x+1,t)}  + d_n \tau^{N(x+1,t)+N(x-1,t)}  \right)  \Big\rangle
 \label{te_mom2}
\end{align}
 with
 \begin{align}
 & a_n =  \frac{q(1-\tau^{-n})(-\tau^3+\tau^n)}{(1-\tau)^2} , \quad 
   b_n   =  \frac{p(1-\tau^{-n})(\tau^2-\tau^n)}{(1-\tau)^2}, \\
  & c_n     =  \frac{q(1-\tau^{-n})(\tau^2-\tau^n)}{(1-\tau)^2} , \quad 
  d_n  =  \frac{p(1-\tau^{-n})(-\tau+\tau^n)}{(1-\tau)^2}.  
 \label{def:abcd}
\end{align}
Denote the rhs of (\ref{te_mom2}) by $V_n(x,t)$. 
For the general case with $(y_i,m_i)$'s, the evolution equation is given as a linear 
combination for the $m_i$-th moments at $y_i$, namely, 
\begin{equation}
 \frac{d}{dt}
 \langle \prod_{i=1}^k \tau^{\sum_i m_i N(y_i,t)} \rangle 
 =
 \sum_{i=1}^k V_{m_j}(y_j,t). 
 \label{te_gen}
\end{equation}
\end{proposition} 

Note that the evolution equation for the moment (\ref{te_mom2}) can be seen as a linear equation with 
inhomogeneous term given by the (lower-order) $\tau$-correlations. If one can find quantities 
which satisfy both (\ref{te_gen}) and  (\ref{te_mom2}), they will
give the formulas for $\tau$-correlations with 
$x_1\leq \cdots \leq x_N$ including $\tau$-moments. 

To prove the proposition, the following simple identity will be useful. 
\begin{equation}
\label{tauN_rel}
\tau \tau^{2N(x-1,t)} + \tau^{2N(x,t)}-(1+\tau)\tau^{N(x-1,t)+N(x,t)} = 0, 
\end{equation}
which is equivalent to $(\tau^{\eta_x}-\tau)\eta_x=0$ and can be
checked easily by observing that the left hand side vanishes 
irrespective of the value of the particle occupation 
at $x$ being $\eta_x = 0$ or 1.

\noindent
{\it Proof of Proposition \ref{prop:eeASEPmoment}. } 
First we prove (\ref{te_mom2}) by a  stochastic calculus similar to that in  subsection \ref{tau-cor}.  
Considering the change of $\tau^{nN(x,t)}$ during the infinitesimal time duration $dt$, we find
\begin{equation}
 \tau^{n dN(x,t)}-1 
 =
 \begin{cases}
  \tau^n-1 , & \text{with rate} \quad q \eta_{x+1}(t) (1-\eta_x(t)), \\
  1/\tau^{n}-1, & \text{with rate} \quad p \eta_x(t) (1-\eta_{x+1}(t)), \\
  0, & \text{otherwise}.
 \end{cases}
\end{equation}
Because $\eta_x$ and $\eta_{x+1}$ are binary variables, there are some coefficients 
$a_n,b_n,c_n,d_n$ such that the following holds. 
\begin{equation}
 q(\tau^n-1)\eta_{x+1}(1-\eta_x) + p(1/\tau^n-1)\eta_x (1-\eta_{x+1})
 =
 a_n + b_n \tau^{-\eta_x} + c_n \tau^{\eta_{x+1}} + d_n \tau^{\eta_{x+1}-\eta_x}. 
\end{equation}
One can readily check that they are exactly given by (\ref{def:abcd}). Then 
(\ref{te_mom2}) follows immediately. 

Next we consider the general case. 
When $\{x_i\}$'s consist of several separate clusters ($y_{i+1}-y_i>1$ for some $i$), it is obvious 
that the time evolution equation is a linear combination of those for each cluster. Hence in the following 
we consider the case of a single cluster, i.e., $y_{i+1}-y_i=1,\forall i$. Then the shape of the cluster is 
determined by a list $(m_1,\cdots, m_k)$ with $\sum_{j=1}^k m_j = n$.  For example, when $N=4$, there are eight 
possible lists, 1111, 211, 121, 112, 22, 31, 13, 4. 
We now order them. On  lists corresponding to different partitions (1111, 211, 22, 31, 4 for $n=4$), we put the 
lexicographic ordering ($1111<211<22<31<4$).  For lists which share the same partition, we put the order by 
the value of $\sum_{j=1}^k j m_j$, e.g., $211<121<112$. When these values are the same, we don't specify 
the order between them. For example 312 and 231 have the same values. 

Then our basic strategy of  proof is to use recursions with respect to this ordering  by applying 
(\ref{tauN_rel}) by writing it as $\tau^{2N(x,t)}=(1+\tau)\tau^{N(x-1,t)+N(x,t)}-\tau \tau^{2N(x-1,t)}$. 
For a given $(m_1,\cdots,m_k)$'s, find $\text{max}~ m_i$  and take the smallest $j$ s.t. $m_j = \text{max}~ m_i$.
If $m_{j-1} \leq m_j-2$, we apply (\ref{tauN_rel}) with $x=y_j$ and rewrite 
$\langle \prod_{i=1}^k \tau^{\sum_i m_i N(y_i,t)} \rangle$  as the sum of two terms with strictly lower order.
For example $224$ can be rewritten as a sum of $233$ and $242$. 
If $\max(1,m_{j-l-1}) < m_{j-k} = m_j-1, k=1,\ldots,l, 1\leq l<j$, where $l$ is the number of $m$'s having the same value $m_j$, 
one applies (\ref{tauN_rel}) with $x=y_{j-l}$. For example $334$ is rewritten as a sum of $1234$ and $2134$. 
The remaining case is when $m_i=1,1\leq i\leq j-1, m_j=2$. When $j\neq k$, we apply (\ref{tauN_rel}) 
with $x=y_j$ and when $j=k$, we apply (\ref{tauN_rel}) by writing it as 
$\tau^{2N(x-1,t)}=(1+\tau^{-1})\tau^{N(x-1,t)+N(x,t)}-\tau^{-1}\tau^{2N(x,t)}$. Anyway we can rewrite 
$\langle \prod_{i=1}^k \tau^{\sum_i m_i N(y_i,t)} \rangle$  as the sum of two terms with lower order.
We can conclude our proof by recalling that $m_i=1,1\leq i\leq k$ case is trivial because this corresponds 
to a $\tau$-correlation for which the time evolution equation was derived in the previous subsection. 
\qed

\medskip
\noindent
{\it Proof of Proposition \ref{prop:ASEPmoment}}.  
From the formulas, it is clear that the expression (\ref{ASEPcor})
 for  $\tau$-correlations and  moments  with $x_1\leq \cdots \leq x_N$ (including 
(\ref{ASEPmoment}) as a special case)  is  consistent with (\ref{tauN_rel}).
Using the same rewriting as in the proof of Proposition  \ref{prop:eeASEPmoment} 
and recursions, one  sees that (\ref{ASEPcor}), (\ref{ASEPmoment}) satisfy (\ref{te_gen}). 
Then from the discussions below (\ref{te_gen}),
it is automatically guaranteed that the formula (\ref{ASEPmoment}) gives the correct formula 
for the $\tau$-moment. 
\qed 
\vskip 0.2cm

\noindent
{\it Remark.} 
The evolution equation for the moments (\ref{te_mom2}) in Proposition \ref{prop:ASEPmoment} for $n\geq 3$ 
can also be derived by applying (\ref{tauN_rel}) and using induction. This alternative proof
will be explained in 
Appendix \ref{AltTem}. 

\medskip
\noindent
{\it Remark.} 
Recently, there appeared a work on the stationary ASEP \cite{Aggarwal2018}, in which  
ASEP is treated as a special case of the higher spin vertex model. In such
higher spin vertex models,
more than one particle  can occupy  the same site. Then,  one can readily
 set $x_i=x,\forall i$ in the correlation found in \cite{Aggarwal2018}
to get a formula for the moment. (See also  \cite{BorodinPetrov}.)

\subsection{Residue expansion of the contour integral}
Before  considering the symmetric limit, it is useful
  to perform an expansion of (\ref{ASEPmoment}) in terms of residues. 
 Using the definition of the contours in Proposition \ref{prop:ASEPcor},
 we write (\ref{ASEPmoment}) as the sum of $2^n$ terms.  
  Each term is an integral with $k$  contours 
 around $-1$ and $n-k$ nested contours around $\tau \th_+$, and is indexed by 
 a subset $P (\subset\{1,\ldots , n\})$ of cardinality $|P|=k$, corresponding to the contours around $-1$. 
 Evaluating the residues at poles related to $\rho_+$
(which are at $\tau^i \theta_+,i\in\mathbb{Z}$) 
from $z_n$ to $z_1$, we obtain the following expansion, 
\begin{align}
\langle \tau^{n N(x,t)} \rangle
&= 
\sum_{k=0}^n (-1)^k \prod_{i=1}^{n-k} e^{\Lambda_i}\prod_{i=n-k+1}^n\left( 1-\frac{r_-}{\tau^i r_+} \right)\notag\\
&\quad \times
\sum_{\substack{P\subset\{1,\ldots , n\}\\ |P|=k}} 
\tau^{v(P)} \int_{-1} \cdots \int_{-1} f_P(z_1,\ldots ,z_k) 
\prod_{i=1}^k dz_i ,
\label{ASEPmomentP}
\end{align}
where $v(P)=||P||-k$ with $||P||=\sum_{i\in P}i$, and  where $\Lambda_i = \Lambda_i(x,t)$ is defined by
\begin{equation}
e^{\Lambda_i} = \left(\frac{1+\tau^i \theta_+}{1+\tau^{i-1}\theta_+}\right)^x e^{\ve(\tau^i \th_+)t} .
\label{Lambdai}
\end{equation}
Notice that because of the order of evaluation of the residues there are no contributions from the 
poles coming from $1/(z_i-\tau z_j)$.  When the pole related to $\rho_+$ is taken for $z_j$, 
the denominator in $F_{x,t}$ is canceled by the numerator in the factor $(z_i-z_j)/(z_i-\tau z_j)$ and 
is replaced by its denominator. This procedure gives an  extra factor of $\tau^{j-1}$, which leads to 
the overall  $\tau$ factor with the exponent $n(n-1)/2-\sum_{j\neq P}(j-1) = \sum_{i\in P}(i-1)=v(P)$.  
The function $f_P$ depends also on $P$ but the dependence will become irrelevant in the symmetric 
limit which is our  main interest, see (\ref{IntJ}). 

To understand the structure, it would be useful to see a few examples.  The above expansion reads, for $n=1$,  
\begin{equation}
\langle \tau^{N(x,t)} \rangle
=
e^{\Lambda_1} - (1-\frac{r_-}{\tau r_+})\int_{-1} F_{x,t}(z) dz
\end{equation}
and for $n=2$, 
\begin{align}
\langle \tau^{2N(x,t)} \rangle
&=
e^{\Lambda_1+\Lambda_2}
-
e^{\Lambda_1}(1-\frac{r_-}{\tau^2 r_+})
\left(  \int_{-1} dz_1 \frac{\left(\frac{1+z_1}{1+z_1/\tau}\right)^{x_1} e^{\ve(z_1)t}}
                                      {(1-\frac{z_1}{\tau^2 \th_+})(z_1-\th_-)}
       +  \tau \int_{-1} dz_2 \frac{\left(\frac{1+z_2}{1+z_2/\tau}\right)^{x_1} e^{\ve(z_2)t}}
                                      {(1-\frac{z_2}{\th_+})(z_2-\th_-)}
                                    \right)   \notag\\
&\quad +
(1-\frac{r_-}{\tau r_+})(1-\frac{r_-}{\tau^2 r_+})\tau
\int_{-1} dz_1 \int_{-1} dz_2 \frac{z_1-z_2}{z_1-\tau z_2} F_{x,t}(z_1)F_{x,t}(z_2).                                    
\label{mom2}
\end{align}

\section{The Symmetric Exclusion Process limit}
\label{sec:symlim}
In this section we take the symmetric limit of the results in the previous section and 
prove our key result in the paper, Proposition \ref{prop:Nmom}. 
Let us denote the height  for SEP by $N=N(x,t)$ and the height 
for ASEP  by $N_\text{ASEP}=N_\text{ASEP}(x,t)$ to make a clear distinction. 
We see, for $\tau=1-\epsilon, \epsilon \to 0$, 
\begin{equation}
\label{SEPlim}
\tau^{N_\text{ASEP}} = 1-\epsilon N + O(\epsilon^2) 
\end{equation}
and hence 
\begin{equation}
 \langle (1-\tau^{N_\text{ASEP}})^n \rangle 
 = 
 \epsilon^n \langle N^n \rangle + o(\epsilon^n) . 
\label{SEPlimn}
\end{equation}
In other words, the $n$th moment of SEP can be found by the lowest $n$th order (in $\epsilon$)
coefficient from the moments of ASEP with degree $n$ and lowers. 
In terms of $\tau$-moments for ASEP,  the l.h.s.  of (\ref{SEPlimn}) is expanded as
\begin{equation}
\langle (1-\tau^{N_\text{ASEP}})^n \rangle 
=\sum_{j=0}^n (-1)^{n-j} \binom{n}{j} \langle \tau^{(n-j)N_\text{ASEP}} \rangle . 
\end{equation}

In the $\tau$-deformed moment formula for ASEP (\ref{ASEPmomentP}), 
the integrand in the $k$-fold integral $f_P$ depends on $P$ and the formula is 
fairly involved. But applying the change of variable (\ref{cov}) in the pole expansion formula 
(\ref{ASEPmomentP}),
and taking the limit, $\epsilon=1-\tau\downarrow 0$, one observes that it depends 
only on $k$ to the leading order as follows, 
\begin{equation}
\label{IntJ}
  \int_{-1} \cdots \int_{-1}  f_P(z_1,\ldots ,z_k)  \prod_{i=1}^k dz_i 
  = 
  \e^k r_+^k J_k + O(\e^{k+1}) 
\end{equation}
where $J_k$ was defined in (\ref{Jk}). 
To understand the appearance of the  common $J_k$, it may be useful to
look at a specific case:  for example,  the  integrands 
of the $z_1$ integrals in the second term on  the right hand side of (\ref{mom2})
are different but become
the same in the $\tau\to 1$ limit. 
The factor $\e^k r_+^k$ is due to the change of variable (\ref{cov}) 
as in the transformation  from (\ref{ASEPcor}) to (\ref{ASEPcor2}).  

Substituting (\ref{IntJ}) in (\ref{ASEPmomentP}), we obtain
\begin{align}
\langle \tau^{n N_\text{ASEP}} \rangle
&=
\sum_{k=0}^n \e^k \prod_{i=1}^{n-k} e^{\Lambda_i}\prod_{i=n-k+1}^n \left(r_+-\frac{r_-}{\tau^i} \right) 
{n \brack k}_\tau (-1)^k \tau^{k(k-1)/2} (J_k+O(\e)) 
\end{align}
where we used $\sum_{P\subset\{1,2,\cdots,n\},|P|=k}\tau^{||P||-k}={n \brack k}_\tau \tau^{k(k-1)/2}$ 
(which is proved by multiplying by $x^k$ and taking a sum $\sum_{k=0}^n$ to get 
a  $q$-binomial theorem $\prod_{k=0}^{n-1}(1+x \tau^k) = \sum_{k=0}^n {n \brack k}_\tau \tau^{k(k-1)/2}x^k.$)  
Hence 
\begin{align}
&\quad \langle (1-\tau^{N_\text{ASEP}})^n \rangle 
= 
\sum_{j=0}^n \binom{n}{j} (-1)^{j} \langle \tau^{j N_\text{ASEP}} \rangle \notag\\
&=
\sum_{j=0}^n \binom{n}{j} (-1)^{j} \sum_{k=0}^j \e^k {j \brack k }_\tau 
\prod_{i=1}^{j-k} e^{\Lambda_i} \prod_{i=j-k+1}^j \left( r_+-\frac{r_-}{\tau^i} \right) (-1)^k \tau^{k(k-1)/2} (J_k+O(\e)) 
\notag\\
&= 
\sum_{k=0}^n 
 \left\{ \sum_{j=k}^n (-1)^{j-k} \binom{n}{j} {j \brack k}_\tau
 \prod_{i=1}^{j-k} e^{\Lambda_i}\prod_{i=j-k+1}^{j} \left( r_+-\frac{r_-}{\tau^i} \right)  \right\} 
 \tau^{k(k-1)/2}  \epsilon^k (J_k+O(\e)) . 
\label{momJ}
\end{align} 

Let us define 
\begin{equation}
\mu_{n,k}(\e) 
= 
\sum_{j=k}^n (-1)^{j-k} \binom{n}{j} {j \brack k}_\tau
 \prod_{i=1}^{j-k} e^{\Lambda_i}\prod_{i=j-k+1}^{j} \left( r_+-\frac{r_-}{\tau^i} \right) .
\end{equation}
We will show that the following  limit exists: 
\begin{equation}
  \tilde{m}_{n,k} = \lim_{\e\to 0} \mu_{n,k}(\e)/\e^{n-k} ,
  \label{limitexists}
\end{equation}
and, therefore, $\mu_{n,k}(\e) = (\tilde{m}_{n,k}+O(\e))\e^{n-k}$.
 Hence, we can neglect $O(\e)$ 
in (3.7) so that we obtain 
\begin{equation}
\langle N^n \rangle = \sum_{k=0}^n \tilde{m}_{n,k}J_k.  
\end{equation}
Comparing  with (1.34), it remains  to show that 
\begin{equation}
  \tilde{m}_{n,k}=m_{n,k}.
  \label{m=mtlde}
  \end{equation}
Let us introduce generating functions of $\mu_{n,k}(\e)$ and $m_{n,k}$ for a fixed 
$k=0,1,\ldots,n$ as 
\begin{align}
 W_k(\l) &= \sum_{n=0}^\i \frac{(\l/\e)^n}{n!} \mu_{n,k}(\e), \\
 M_k(\l) &= \sum_{n=0}^\i \frac{\l^n}{n!} \tilde{m}_{n,k}. 
\end{align}

The statements (\ref{limitexists}) and (\ref{m=mtlde}) are proved
 by the following proposition:

\begin{proposition}
\label{prop:Mkt}
\begin{equation}
 M_k(\l) = \lim_{\e\to 0} \e^k W_k(\l) =  \frac{\o^k}{k!} (1+\rho_+(e^{\l}-1))^x .
 \label{Mkl}
\end{equation}
\end{proposition} 

\noindent 
{\it Proof.} 
From  equation (3.8) for $k=0$, we obtain
\begin{equation}
\label{G0X}
 W_0(\l) = e^{\l/\e}  \sum_{n=0}^\i \frac{(-\l/\e)^n}{n!} \prod_{i=1}^n e^{\Lambda_i} . 
 \end{equation}
Here $\Lambda_i$ is given by (\ref{Lambdai}). First let us consider the $t=0$ case, for which  
this expression can be transformed as follows
\begin{align}
 &\quad 
 e^{\l/\e} \sum_{n=0}^\i \frac{(-\l/\e)^n}{n!}\left(\frac{1+\tau^n \th_+}{1+\th_+}\right)^{x} \notag\\
 &=
 e^{\l/\e} \left(\frac{1}{(1+\th_+)^x}\sum_{n=0}^\i \frac{(-\l/\e)^n}{n!}\sum_{m=0}^x \binom{x}{m} (\tau^n\th_+)^m \right)\notag\\
 &=
\frac{1}{(1+\th_+)^x}\sum_{m=0}^x \binom{x}{m} \th_+^m e^{-(\tau^m-1)\l/\e}. 
\label{Mkt}
\end{align}
 Using   
$  e^{-(\tau^{m}-1)\l/\e} \to e^{m\l}$ as $\e\to 0$, we conclude 
 from  (\ref{Mkt}) that $W_0(\l)$
converges to 
$M_0(\l) = (1+\rho_+(e^{\l}-1))^x$  in the small  $\e$ limit. 

When $t>0$, (\ref{Mkt}) should be multiplied by $e^{\sum_{i=1}^n\epsilon(\tau^i \th_+)t}$. 
We will argue that this factor does not change $\lim_{\e\to 0} W_0(\l)$. 
First we rewrite $\g(z)$ in (\ref{gamma}) as $\g(z) = q(1-\tau)(1/(1+z/\tau)-1/(1+z))$
and find 
\begin{equation}
 e^{\sum_{i=1}^n \g(\tau^i \th_+)t}
 =
 e^{ \frac{q(1-\tau)}{1+\th_+}t} e^{-\frac{q(1-\tau)}{1+\tau^n \th_+}t}
 =
 e^{ \frac{q \e}{1+\th_+}t} \times \sum_{j=0}^{\infty} \frac{(qt)^j}{j!(1+\tau^n \th_+)^j} \e^j .
\end{equation}
The first factor $e^{ \frac{q \e}{1+\th_+}t}$ is independent of $n$ and tends to unity in 
the $\e\to 0$ limit so that obviously it does not influence the limit $\lim_{\e\to 0}  W_0(\l)$.
For the second factor, the $j$th term cancels the factor $(1+\tau^n \th_+)^x$ in the leftmost 
side of (\ref{Mkt}) and hence we have 
\begin{align}
&\quad 
 e^{\l/\e} \sum_{n=0}^\i \frac{(-\l/\e)^n}{n!}\left(\frac{1+\tau^n \th_+}{1+\th_+}\right)^{x} e^{-\frac{q(1-\tau)}{1+\tau^n \th_+}t}\notag\\
 &=
 \sum_{j=0}^{\infty} \frac{1}{j!} \left(\frac{qt}{1+\th_+}\right)^j  \left(\frac{1}{1+\rho_+(e^{\l}-1)}\right)^j \e^j \times M_0(\l)|_{t=0}\notag\\
 &=
 \exp \left(  \frac{qt}{(1+\tau^n \th_+)(1+\rho_+(e^{\l}-1))} \e  \right) M_0(\l)|_{t=0} .
\end{align}
Here in the first equality we used (\ref{Mkt}). 
With this expression it is easy to see that, when  $\e\to 0$,
we have  $\lim_{\e\to 0} W_0(\l) = M_0(\l)$ for all $ t \ge 0$.

Using similar calculations for $k >0$ we obtain
\begin{align}
 \e^k W_k(\l) 
 &=
 e^{\l/\e}\frac{\l^k}{[k]_\tau!}
 \sum_{l=0}^{\infty} (-1)^l \frac{[l+k]_\tau! (\l/\e)^l}{[l]_\tau! (l+k)!}
 \prod_{i=l+1}^{l+k}(r_+-r_-/\tau^i) \prod_{p=1}^l e^{\Lambda_p}\notag\\
 &=
 e^{\l/\e}\frac{\e^k}{[k]_\tau!}
 \sum_{l=0}^{\infty} \frac{[l]_\tau! (-\l/\e)^l}{[l-k]_\tau! l!}
 \prod_{i=l-k+1}^{l}(r_+-r_-/\tau^i) \prod_{p=1}^{l-k} e^{\Lambda_p}.
\end{align}
Note that in the middle expression the sum over $l$ can be made starting from 
$-k$ due to the factor $[l+k]_{\tau}!/[l]_\tau!$. Accordingly we made a shift $l\to l-k$
in the second equality. 

Expanding the two factors $[l+k]_{\tau}!/[l]_\tau! = \prod_{i=0}^{k-1} (1-\tau^{l-i})$ and
$\prod_{i=l-k+1}^{l}(r_+-r_-/\tau^i)$, one can perform the sum over $l$ and do 
similar calculations as for the $k=0$ case above. Observing that an extra factor 
$\tau^i$ in front of $\tau^l$ would not make any difference
in the $\e\to 0$ limit, we can now replace the
factor $[l+k]_{\tau}!/[l]_\tau!$ by $(1-\tau)^l$, $\prod_{i=l-k+1}^{l}(r_+-r_-/\tau^i)$ by 
$(r_+-r_-/\tau^l)^k$ and $\prod_{i=1}^{l-k} e^{\Lambda_i}$ by $\prod_{i=1}^{l} e^{\Lambda_i}$
with errors of $O(\e)$ in the final result. Therefore expanding the first two factors with binomial 
coefficients we find
\begin{align}
 \e^k W_k(\l) 
 &=
\frac{e^{\l/\e}}{k!} 
 \sum_{j=0}^k (-1)^j \binom{k}{j} \sum_{i=0}^k \binom{k}{i} r_+^{k-i} r_-^i 
   \sum_{l=0}^{\infty} (-1)^l \frac{ (\l/\e)^l}{ l!}\tau^{(j-i)l}
 \prod_{p=1}^l e^{\Lambda_p} +O(\e). 
\end{align}
When $t=0$ one applies the same calculation as in (\ref{Mkt}) to find 
\begin{align}
e^{\l/\e}\sum_{l=0}^{\infty} (-1)^l \frac{ (\l/\e)^l}{ l!}\tau^{(j-i)l}
 \prod_{p=1}^l e^{\Lambda_p}
 =
 \frac{1}{(1+\th_+)^x}\sum_{m=0}^x \binom{x}{m} \th_+^m e^{-(\tau^{m+j-i}-1)\l/\e} 
\xrightarrow{\e\to 0}
 e^{(j-i)\l} W_0(\l).
 \end{align}
 One can also check that $t>0$ would not change the limit as in the $k=0$ case. 
 Observing 
 \begin{equation}
 \sum_{j=0}^k (-1)^j \binom{k}{j} \sum_{i=0}^k \binom{k}{i} r_+^{k-i} r_-^i e^{(j-i)\l}
 =
 (e^{\l}-1)^k (r_+ + r_-e^{-\l})^k = \o^k, 
 \end{equation} 
 we finally arrive at (\ref{Mkl}). 

\noindent
Comparing (\ref{Mkl}) with (\ref{Mk}), one finds $\tilde{m}_{n,k}=m_{n,k}$, concluding 
the proof of Proposition \ref{prop:Mkt}. 
\qed

\section{Fredholm determinant for the integrals $J_n$}
\label{sec:JFdet}
  In this section we prove Proposition \ref{prop:JFdet} by adapting the arguments in,  
  for example,  \cite{TW2008b,DG2009}. 
  Let us first recall 
\begin{align}
  J_n 
 &= 
 \int_{C_0} \cdots \int_{C_0} \prod_{1\leq i<j\leq n}\frac{\xi_i-\xi_j}{\xi_i\xi_j+1-2\xi_j}
 \prod_{i=1}^n 
\frac
{\xi_i^x e^{\ve(\xi_i)t} d\xi_i}
{(1-\xi_i)^2}
\end{align}
where $C_0$ is a contour around the origin with radius so small that
the poles from the factor $1/(\xi_i\xi_j+1-2\xi_j)$ in the integrand are not included. 
We rewrite $J_n$ as
\begin{align}
 J_n 
&= 
 \int_{C_0} \cdots \int_{C_0} 
\prod_{1\leq i\neq j\leq n}\frac{1}{\xi_i\xi_j+1-2\xi_j}
\cdot \prod_{1\leq i<j\leq n} (\xi_i\xi_j+1-2\xi_i)
\prod_{1\leq i<j\leq n}(\xi_i-\xi_j)
\cdot 
\prod_{i=1}^n 
\frac
{\xi_i^x e^{\ve(\xi_i)t} d\xi_i}
{(1-\xi_i)^2} .
\end{align}
We can relabel the indices by $i$ by an arbitrary permutation ${\sigma\in S_n}$,
 where $S_n$  is the permutation group  $(1,2,\cdots,n)$. Taking into account
 that the products over all ${1\leq i\neq j\leq n}$ are  permutation-symmetric  and 
 the Vandermonde product is antisymmetric, the above  integral can be rewritten as
\begin{align}
&
 \int_{C_0} \cdots \int_{C_0} 
\prod_{1\leq i\neq j\leq n}\frac{1}{\xi_i\xi_j+1-2\xi_j}
\cdot \sgn\sigma \prod_{1\leq i<j\leq n}  (\xi_{\sigma(i)}\xi_{\sigma(j)}+1-2\xi_{\sigma (i)})
 \notag\\
 &\times
 \prod_{1\leq i<j\leq n}(\xi_i-\xi_j) 
 \prod_{i=1}^n 
\frac
{\xi_i^x e^{\ve(\xi_i)t} d\xi_i}
{(1-\xi_i)^2} 
\notag
\end{align}
Summing over all permutations $\sigma$ in the set $S_n$  (and normalizing by $1/n!$), we have 
\begin{align}
J_n  &= 
 \int_{C_0} \cdots \int_{C_0} 
\prod_{1\leq i\neq j\leq n}\frac{1}{\xi_i\xi_j+1-2\xi_j}
\cdot
\prod_{1\leq i<j\leq n}(\xi_i-\xi_j)
\cdot 
\prod_{i=1}^n 
\frac
{\xi_i^x e^{\ve(\xi_i)t} d\xi_i}
{(1-\xi_i)^2} 
\notag
\\
&\hspace{2cm}
\times\frac{1}{n!}
\sum_{\sigma\in S_n} \sgn\sigma \prod_{1\leq i<j\leq n}
(\xi_{\sigma(i)}\xi_{\sigma(j)}+1-2\xi_{\sigma (i)})
\notag
\\
&=
 \int_{C_0} \cdots \int_{C_0} 
\prod_{1\leq i\neq j\leq n}\frac{\xi_i-\xi_j}{\xi_i\xi_j+1-2\xi_j}
\cdot
\prod_{1\leq i<j\leq n}(\xi_i-\xi_j)
\cdot 
\prod_{i=1}^n 
\frac
{\xi_i^x e^{\ve(\xi_i)t} d\xi_i}
{(1-\xi_i)^2} 
\notag
\\
 &=
\int_{C_0} \cdots \int_{C_0} \det(K_{x,t}(\xi_i,\xi_j))_{i,j=1}^n \prod_{i=1}^n d\xi_i,
\label{b2}
\end{align}
where  
\begin{equation}
 K_{x,t}(\xi_1,\xi_2) = \frac{\xi_1^x e^{\ve(\xi_1) t}}{\xi_1\xi_2+1-2\xi_2}
\end{equation}
is the kernel that appears in Theorem \ref{thm:GF}. 
The equation \eqref{b2} says that  $J_n$ is the $n$th term in the Fredholm expansion 
of the Fredholm determinant given in (\ref{Jgen}). 

In the derivation of (\ref{b2}), we use the following two combinatorial identities, 
\begin{align}
&\sum_{\sigma\in S_n}
\sgn\sigma
\prod_{1\leq i<j\leq n}
(\xi_{\sigma(i)}\xi_{\sigma(j)}+1-2\xi_{\sigma (i)})
=n!\prod_{1\leq i<j\leq n}(\xi_j-\xi_i),
\label{b4}
\\
&\prod_{k=1}^n
\frac{1}{(1-\xi_k)^2}
 \prod_{i\neq j}\frac{\xi_i-\xi_j}{\xi_i\xi_j+1-2\xi_j}
=
\det\left(\frac{1}{\xi_i \xi_j+1-2\xi_i}\right)_{i,j=1,\cdots,n} . 
\label{b5}
\end{align}
The last equation  has been proved in \cite{TW2008b}. The identity~\eqref{b4} can be shown as follows:
Changing the variable as $t_i=1/(\xi_i-1)$, \eqref{b4} becomes
\begin{align}
\sum_{\sigma\in S_n}
\sgn\sigma
\prod_{1\leq i<j\leq n}
(1+t_{\sigma(i)}-t_{\sigma(j)})
=n!\prod_{1\leq i<j\leq n}(t_i-t_j).
\label{b4d}
\end{align}
We introduce an arbitrary parameter $a\in\C$ and prove a slightly more general formula,
\begin{align}
\sum_{\sigma\in S_n}
\sgn\sigma
\prod_{1\leq i<j\leq n}
(a+t_{\sigma(i)}-t_{\sigma(j)})
=n!\prod_{1\leq i<j\leq n}(t_i-t_j).
\label{b4d2}
\end{align}
 Expanding the left hand side  with respect to $a$, we observe that 
the coefficient of $a^k$ will be a linear combination of  terms of the form 
\begin{align}
\sum_{\sigma\in S_n}
\sgn\sigma \, 
t_{\sigma(1)}^{m_1}t_{\sigma(2)}^{m_2}\cdots t_{\sigma(n)}^{m_n}
\label{b8}
\end{align}
where $m_j\in\Z_{\ge 0}~$, for $j=1,2,\cdots,n$,  satisfy the conditions 
$0\le m_j\le n-1$ and $~\sum_{j=1}^n m_j=\frac{n(n-1)}{2}-k.$
Then,  for   $k\ge 1$,  at least some pair of $m_j,~j=1,2,\cdots,n$  
should have the same value: this implies that~\eqref{b8} vanishes.
 The polynomial \eqref{b4d2} does not depend on $a$ and we have 
\begin{align}
\sum_{\sigma\in S_n}
\sgn\sigma
\prod_{1\leq i<j\leq n}
(a+t_{\sigma(i)}-t_{\sigma(j)})
=
\sum_{\sigma\in S_n}
\sgn\sigma
\prod_{1\leq i<j\leq n}
(t_{\sigma(i)}-t_{\sigma(j)})
=n!\prod_{1\leq i<j\leq n}(t_i-t_j).
\end{align}
\qed


\section*{Acknowledgments}
The authors would like to thank P. Krapivsky, S. Olla, L. Petrov,  S. Sethuraman, H. Spohn for useful remarks. They  are also grateful  to S. Mallick  for a careful  reading of the manuscript.
We thank the JSPS core-to-core program "Non-equilibrium dynamics of soft matter and information"  
which initiated this work. Parts of this work were performed during stays at ICTS  Bangalore and 
at KITP  Santa Barbara. 
This research was supported in part by the National Science Foundation under 
Grant No. NSF PHY11-25915. The works of T.I and T.S. are also supported by JSPS 
KAKENHI Grant Numbers JP25800215, JP16K05192 and 
JP13321132, JP15K05203, JP16H06338, 18H01141, 18H03672 respectively. 

%
%
\appendix
\section{Moments and cumulants}
\label{sec:momcum}
In this appendix, we summarize a few basic formulas about the moments, cumulants and their generating functions, 
which are useful for  discussions in the main text. 

For a random variable $X$, the moment is 
\begin{equation}
 m_n = \langle X^n \rangle, ~~ n\in\N .
\end{equation}
Note $m_0=1$ for any $X$. 
The moment generating function is defined by 
\begin{equation}
\label{mgen}
 M(\lambda) = \sum_{n=0}^{\infty} \frac{\lambda^n}{n!} m_n . 
\end{equation}
The cumulant generating function $K(\lambda)$ and the cumulant $c_n$ are 
defined through 
\begin{equation}
\label{cgen}
 K(\lambda) = \log M(\lambda) = \sum_{n=1}^{\infty} \frac{\lambda^n}{n!} c_n . 
\end{equation}
By convention we set $c_0=0$. 

The relations between the moments and cumulants can be written down explicitly. 
The moment can be written in terms of cumulants as, for $n\in\Z_+$,  
\begin{equation}
\label{momcum}
 m_n = \sum_{\substack{\nu \vdash n \\ \nu=1^{l_1} 2^{l_2} \cdots }}
            a_{\nu}^n c_1^{l_1} c_2^{l_2} \cdots 
\end{equation}
with 
\begin{equation}
 a_{\nu}^n = n! \prod_{j=1}^n \frac{1}{l_j!}\left(\frac{1}{j!}\right)^{l_j} .
 \label{anun}
\end{equation}
For example, 
\begin{align}
 m_1 &= c_1, \\
 m_2 &= c_1^2+c_2, \\
 m_3 &= c_1^3+3c_1 c_2 + c_3, \\
 m_4 &= c_1^4+6 c_1^2 c_2 + 3 c_2^2 + 4c_1 c_3 + c_4, \\
 m_5 &= c_1^5+10 c_1^3 c_2 + 15 c_1 c_2^2 + 10c_1^2 c_3  + 10 c_2 c_3 + 5 c_1 c_4 + c_5 .
\end{align}
Conversely the cumulant is written in terms of moments as
\begin{equation}
 c_n = \sum_{\substack{\nu \vdash n \\ \nu=1^{l_1} 2^{l_2} \cdots }}
           \tilde{a}_{\nu}^n m_1^{l_1} m_2^{l_2} \cdots 
\end{equation}
with 
\begin{equation}
 \tilde{a}_{\nu}^n = (-1)^{l-1} (l-1)! a_{\nu}^n, \quad l=l_1+l_2+\cdots .
 \label{anunt}
\end{equation}

The first few examples are given by 
\begin{align}
 c_1 &= m_1, \\
 c_2 &= m_2-m_1^2, \\
 c_3 &= m_3-3m_1 m_2+2m_1^3, \\
 c_4 &= m_4-4m_3m_1-3m_2^2+12m_2 m_1^3-6m_1^4, \\
 c_5 &= m_5-5m_4m_1-10m_3m_2+20m_3 m_1^2+30m_2^2 m_1 -60 m_2 m_1^3+24m_1^5 .
\end{align}

%

\vskip 0.3cm

Now let us set  
\begin{align}
 \alpha_{n,l}(a,b)  
 = 
\sum_{\substack{\nu \vdash n \\ \nu=1^{l_1} 2^{l_2} \cdots \\ l_1+l_2+\cdots = l} }
 \frac{n!}{ \prod_{j=1}^n l_j!} \,  \prod_{j=1}^n 
\left(\frac{a + (-1)^jb }{j!}\right)^{l_j}.
\end{align} 
This is the same formula
as the one defined in (\ref{def:alphanl}). By the general relations between the 
moments and cumulants explained above, this is useful for example in the description of  cumulants 
when the moments are known. 

\smallskip
\noindent
{\it Example 1}. When the moment generating function is $M(\l) = 1+a(e^{\l}-1)$, the moments are
$m_0=1$ and $m_n=a$ for $n\in\mathbb{Z}_+$. The cumulant generating function is 
$K(\l)=\log(1+a(e^{\l}-1))$ and the cumulants are 
\begin{align}
 c_n 
 &=  
 \sum_{\substack{\nu \vdash n \\ \nu=1^{l_1} 2^{l_2} \cdots }}
           \tilde{a}_{\nu}^n a^{\sum_{j=1}^{\infty}l_j}
 = 
 \sum_{l=0}^n (-1)^l(l-1)! 
 \sum_{\substack{\nu \vdash n \\ \nu=1^{l_1} 2^{l_2} \cdots \\ l_1+l_2+\cdots = l} }
 \frac{n!}{ \prod_{j=1}^n l_j!} \,  \prod_{j=1}^n 
\left(\frac{a}{j!}\right)^{l_j} \notag\\
 &=
 \sum_{l=0}^n (-1)^l (l-1)! \a_{n,l}(1,0) a^l. 
\end{align} 

\smallskip
\noindent
{\it Example  2}. 
 When the moment generating function is $M(\l) = 1+a(e^{\l}-1)+b(e^{-\l}-1)$, the momenta are
 $m_0=1$,  $m_n=a+b$ for $n=2j$ and $m_n=a-b$ for $n=2j-1$ 
   with $j\in\mathbb{Z}_+$. The cumulant generating function is 
$K(\l)=\log(1+a(e^{\l}-1)+b(e^{-\l}-1))$ and the cumulants are 
\begin{equation}
 c_n = \sum_{l=0}^n (-1)^l (l-1)! \a_{n,l}(a,b). 
\end{equation}

\section{Asymptotics of $I_n$}
\label{sec:asym} 
We prove Proposition \ref{prop:Iasym}. Our arguments are a generalization of those in \cite{DG2009}, 
but we use the steepest decent method, rather than applying saddle point methods in two steps. 
In the following the symbol $\sim$ means that the ratio of the left and right hand sides goes to unity 
as $t\to\infty$ as in  Proposition \ref{prop:Iasym}. 
For  $x\ge 0$, the  integral \eqref{def:In} can be written as 
\begin{align}
 I_n  
 &= 
 \int_{C_0} \cdots\int_{C_0}\prod_{j=1}^n dz_j z_j^x
\frac{e^{(z_{j+1}+1/z_{j}-2)t}}{z_jz_{j+1}+1-2z_j}
\notag
=\int_{C_0} \cdots\int_{C_0}\prod_{j=1}^ndz_j z_j^x
\frac{e^{(z_{j+1}+1/z_{j}-2)t}-1}{z_jz_{j+1}+1-2z_j}
\\
&=\int_{C_0} \cdots \int_{C_0}\prod_{j=1}^n\frac{dz_j}{z_j}
 z_j^x \int_0^t dt_j
e^{(z_{j+1}+1/z_{j}-2)t_j} 
=
\int_0^t  \cdots 
\int_0^t 
\prod_{k=1}^n  dt_k   
\left[
\prod_{k=1}^n \int_{C_0} 
\frac{d z_k}{z_k}
 e^{t_{k-1}z_k+\frac{t_k}{z_k}-2t_k}
z_k^x \right] \notag\\
&=
\int_0^1 \cdots  \int_0^1 
\prod_{k=1}^n  du_k
2tu_k
 e^{-2tu_k^2}
 \left[
\prod_{k=1}^n \int_{C_0} 
\frac{d w_k}{w_k}
e^{t u_{k-1}u_k(w_k+1/w_k)}
w_k^x \right] \, . 
\label{b18a}
\end{align}
where in the first expression we set $z_{n+1}=z_1$ and in the second equality we 
used $x\geq 0$ so that there is no pole at the origin and also the fact that the radius of the 
contour $C_0$ is so small that there is no pole from the denominators of the integrand. 
In the last equality we 
changed the variables from $t_k,z_k$ to $u_k,w_k$ by 
$t_k=t u_k^2$, $z_k=w_k u_k/u_{k-1}$ (with $u_0=u_n$).

 We now perform a steepest descent analysis on this integral. A standard reference for this 
 method is \cite{Wong1989}
  (for a similar concrete example in the case of a one-dimensional integral see  \cite{Sasamoto1999},
 section  5). By changing the variables
$(u_1,\ldots,u_n) \rightarrow (u,v_1,\ldots,v_{n-1})$ and
 $(w_1 ,\ldots,w_n)  \rightarrow (\theta_1,\ldots, \theta_n)$ as follows
\begin{align}
 u = u_1, \quad \hbox{ and }
 \quad   v_i &= \sqrt{t} (u_{i+1} -  u_i)
 \,\,  \hbox{ for  } \,\,   1 \le i \le n-1 \\
 w_k &= e^{i \theta_k} \,\,  \hbox{ for  } \,\,  1 \le k \le n
\end{align}
 the previous integrals becomes
\begin{align}
I_n  &=2^nt^{\frac{n+1}{2}} \int_0^1 u  du  \left(  \prod_{i=1}^{n-1}
 \int_{-\sqrt{t} u -V_{i-1} }^{\sqrt{t} (1-u) -V_{i-1}} \right)
  e^{-  V_{n-1}^2}   \prod_{i=1}^{n-1}(u + \frac{V_i}{\sqrt{t}}) e^{-v_i^2} dv_i     
\nonumber \\
&\quad\times\prod_{i=1}^{n} \int_{-\pi}^\pi \frac{d \theta_i}{2 \pi}
e^{2t\sum_{k=1}^n( \cos \theta_k -1)  (u + \frac{V_{k-2}}{\sqrt{t}})
 (u + \frac{V_{k-1}}{\sqrt{t}}) + i x \sum_k \theta_k}
\end{align}
where we use $V_{i}= \sum_{j=1}^{i}v_j$ (with $V_0=0$ and $V_{-1}=V_{n-1}$). The integrals over the 
$d \theta_i$'s are transformed further  by the change of variable
\begin{align}
 y_k^2 = 2t( 1 -  \cos \theta_k)
 \end{align}
(Note that when $\theta_k \in [-\pi,\pi]$ the variable  $y_k$
 runs over $[-2\sqrt{t}, 2 \sqrt{t}]$.)
 We obtain, recalling that $x = -2 \sqrt{t} \xi$, 
\begin{align}
I_n  &=2^nt^{\frac{n+1}{2}} \int_0^1 u  du  
\left(  \prod_{i=1}^{n-1} \int_{-\sqrt{t} u -V_{i-1} }^{\sqrt{t} (1-u) -V_{i-1}} \right) 
  e^{-  (\sum_{i=1}^{n-1}v_i)^2}   \prod_{i=1}^{n-1} (u + \frac{V_i}{\sqrt{t}}) e^{-v_i^2} dv_i    
  \nonumber \\
&\quad\times\prod_{i=1}^{n} \left(\int_{-2 \sqrt{t}}^{2 \sqrt{t}}
 \frac{d y_i}{2 \pi \sqrt{t} \sqrt{1 - \frac{y_i^2}{4t}}}
e^{-  y_i^2 (u + \frac{V_{i-2}}{\sqrt{t}})
 (u + \frac{V_{i-1}}{\sqrt{t}}) -2 i\sqrt{t} \xi 
  \arccos(1 - \frac{y_i^2}{2t}) } \right) 
\end{align}
We can now take the limit $t\to \infty$ in this previous integral:
\begin{align}
I_n  &\sim\sqrt{t} \pi^{-n/2} \int_0^1 u^n  du  \left(  \prod_{i=1}^{n-1}
 \int_{-\infty }^{+\infty}  e^{-v_i^2} dv_i  \right)   
 e^{- (\sum_i v_i)^2} 
&\prod_{i=1}^{n} \left(\int_{-\infty}^{\infty} d y_i
e^{-  y_i^2 u^2  -2 i  \xi y_i } \right) 
\end{align}
Evaluating the Gaussian integrals, we   conclude that 
\begin{align}
I_n
\sim
\sqrt{
\frac{t}{n\pi}
} 
\int_0^1
du
e^{-\frac{n \xi^2}{u^2}}=\sqrt{t}\Xi_n (|\xi|) = \sqrt{t}\Xi_n (-\xi)  \,, 
\end{align}
 where the  last equality is found  by changing the variable $u \to 1/u$ and 
 by using the last expression of $\Xi(\xi)$ in (\ref{def:Xi}). 
This concludes the proof of Proposition \ref{prop:Iasym}

\section{Alternative proof of (\ref{te_mom2})}
\label{AltTem}
 We define  $G(x_1,\cdots,x_n)=\langle \tau^{\sum_{j=1}^n N(x_j,t)} \rangle$.
 From  (\ref{tauN_rel}), we first rewrite 
\begin{equation}
G(x,\cdots ,x) = (1+\tau)G(x-1,x,\cdots,x) +\tau G(x-1,x-1,x,\cdots ,x).
\end{equation}
 Using the induction assumption, one can write down the evolution equation for each 
term on the right hand side. For example for the first term we have 
\begin{align}
 &\quad
 \partial_t G(x-1,\underbrace{x,\cdots,x}_{n-1}) \notag\\
 &=
 q G(x-2,\underbrace{x,\cdots,x}_{n-1}) + p G(\underbrace{x,\cdots,x}_n) -(p+q) G(x-1,\underbrace{x,\cdots,x}_{n-1}) \notag\\
 &\quad +
 \frac{q(1-\tau^{-n+1})(-\tau^3+\tau^{n-1})}{(1-\tau)^2} G(x-1,\underbrace{x,\cdots,x}_{n-1}) \notag\\
 &\quad + 
 \frac{p(1-\tau^{-n+1})(\tau^2-\tau^{n-1})}{(1-\tau)^2} G(x-1,x-1,\underbrace{x,\cdots,x}_{n-2}) \notag\\
 &\quad +
 \frac{q(1-\tau^{-n+1})(\tau^2-\tau^{n-1})}{(1-\tau)^2} G(x-1,\underbrace{x,\cdots,x}_{n-2},x+1) \notag\\
 &\quad +
 \frac{p(1-\tau^{-n+1})(-\tau+\tau^{n-1})}{(1-\tau)^2} G(x-1,x-1,\underbrace{x,\cdots,x}_{n-3},x+1). 
\end{align}
By adding a similar equation for the second term, one can write down the evolution equation for the 
left hand side. Using (\ref{tauN_rel}) a few times again as 
$\tau \tau^{2N(x-1,t)} = (1+\tau)\tau^{N(x-1,t)+N(x,t)} -\tau^{2N(x,t)}$, 
one can rewrite the right hand side of the evolution equation 
as a linear combination of $G(\underbrace{x,\cdots,x}_n), G(x-1,\underbrace{x,\cdots,x}_{n-1}), G(\underbrace{x,\cdots,x}_{n-1},x+1),
G(x-1,\underbrace{x,\cdots,x}_{n-2},x+1)$ and check that their coefficients are given by (\ref{def:abcd}). 
This ends the alternative proof of  (\ref{te_mom2}).  
\qed


\end{document}